\documentclass[
reprint,
superscriptaddress,
noeprint,
amsmath,amssymb,
aps,
prx,
]{revtex4-2}

\usepackage[utf8]{inputenc}
\usepackage{graphicx}
\usepackage[dvipsnames]{xcolor}

\usepackage{braket}

\usepackage[colorlinks=true,bookmarks=false,linkcolor=NavyBlue,urlcolor=NavyBlue,citecolor=NavyBlue,breaklinks]{hyperref}

\newcommand{\I}{\mathrm{i}}
\newcommand{\HG}{H\!G}

\begin{document}

\title{Semiparametric estimation in Hong-Ou-Mandel interferometry}
\date{\today}

\author{Valeria Cimini}
\affiliation{Dipartimento di Scienze, Universit\'{a} degli Studi Roma Tre, Via della Vasca Navale, 84, 00146 Rome, Italy}
\affiliation{Dipartimento di Fisica, Sapienza Universit\'{a} di Roma, Piazzale Aldo Moro 5, I-00185 Roma, Italy}

\author{Francesco Albarelli}
\affiliation{Faculty of Physics, University of Warsaw, Pasteura 5, PL-02-093 Warszawa, Poland}
\affiliation{Department of Physics, University of Warwick, Coventry, United Kingdom}

\author{Ilaria Gianani}
\email{ilaria.gianani@uniroma3.it}
\affiliation{Dipartimento di Scienze, Universit\'{a} degli Studi Roma Tre, Via della Vasca Navale, 84, 00146 Rome, Italy}

\author{Marco Barbieri}
\affiliation{Dipartimento di Scienze, Universit\'{a} degli Studi Roma Tre, Via della Vasca Navale, 84, 00146 Rome, Italy}
\affiliation{Istituto Nazionale di Ottica - CNR, Largo Enrico Fermi 6, 50125 Florence, Italy}

\begin{abstract}
We apply the theory of semiparametric estimation to a Hong-Ou-Mandel interference experiment with a spectrally entangled two-photon state generated by spontaneous parametric downconversion.
Thanks to the semiparametric approach we can evaluate the Cramér-Rao bound and find an optimal estimator for a particular parameter of interest without assuming perfect knowledge of the two-photon wave function, formally treated as an infinity of nuisance parameters.
In particular, we focus on the estimation of the Hermite-Gauss components of the marginal symmetrised wavefunction, whose Fourier transform governs the shape of the temporal coincidence profile.
We show that negativity of these components is an entanglement witness of the two-photon state.
\end{abstract}

\maketitle

\section{Introduction}

Two-photon Hong-Ou-Mandel interference~\cite{PhysRevLett.59.2044,branczyk2017hongoumandel,Bouchard_2021} is the key effect that enables many quantum technologies based on photons and their manipulation~\cite{klm,cnotexperiment,PhysRevLett.89.137901,PhysRevLett.93.020504,PhysRevLett.91.083601,Lyons2018,PhysRevLett.93.170501,PhysRevLett.92.047901}.
Its distinctive coincidence dip profile is a signature of the bosonic nature of the photons, and remarkably, its characteristic length is dictated by the two-photon wavepacket, not by their wavelength, ensuring stable and reliable operation, even with modest control of the path lengths, thus allowing for extensions to the multiphoton case~\cite{Nagata726,PhysRevLett.118.153602,PhysRevLett.118.153603,aaronson2010computational,Broome794,Bentivegnae1400255,Stobinskaeaau9674,Zhange1501165}. 

This feature derives from a non-trivial dependence of the interferometric signal on the two-photon spectral wavefunction, however, since it ultimately relies on a symmetrisation operation~\cite{branczyk2017hongoumandel,PhysRevLett.115.193602}, it is an excellent test bed for verifying the degree of indistinguishability and spectral purity of two independent single-photon wavepackets~\cite{PhysRevLett.84.5304,PhysRevLett.100.133601, Santori2002,PhysRevLett.102.123603,tanzilli,PhysRevLett.126.063602,Senellart2017,Francesconi:20,technologies4030025,Dorfman2014,Hua:21,PhysRevLett.93.070503,Patel2010,Lipka2021a}.  
However complicated, this dependence can be inverted to obtain the wavefunction in the experiment, but this requires multiple Hong-Ou-Mandel (HOM) profile acquisitions~\cite{PhysRevLett.115.193602}.
With a single acquisition, some information can nevertheless be extracted, however, in a limited amount~\cite{Fedrizzi_2009,scirepHOM,PhysRevApplied.12.054029}; this can still be an appropriate regime for estimating specific quantities relevant to the wavefunction.

Applying parameter estimation to such cases benefits from a generous pinch of salt when it comes to spelling out the statement of the problem.
Even if we wish to isolate one particular parameter of the wavefunction, e.g. one of its moments, the estimation will unavoidably depend on the whole function, thus requiring, in principle, infinitely many other parameters for its full description. 
This apparently unsolvable problem has an elegant and efficient solution in semiparametric estimation~\cite{bickel1998efficient,Tsiatis2006}.

In standard parameter estimation one has to fix a statistical model, assuming a known dependence of the wavefunction from a finite number of parameters.
On the contrary, the theory of semiparametric estimation deals with models with an infinity of degrees of freedom.
The goal is to extract information about a finite number of parameters of interest, making as few assumptions as possible on the underlying model.
A prototypical example is the estimation of the mean of an unknown probability distribution with finite variance~\cite{Tsiatis2006}.
In the context of quantum technologies, semiparametric methods have recently been applied to superresolution imaging~\cite{PhysRevResearch.1.033006,Tsang2020b} and a fully quantum generalization of the theory has been derived~\cite{PhysRevX.10.031023}.
The related task of estimating a small subset of a finite number of parameters, treating the others as a nuisance, has also been recently studied in the context of quantum estimation theory~\cite{Suzuki2019a,Suzuki2019}.

In this article we apply semiparametric methods to estimating quantities pertinent to the frequency domain, based on the time profile of the coincidence dip in HOM interferometry.
We show that certain quantities, e.g. certain raw moments, cannot be successfully estimated, due to the Fourier transform needed to convert between the two domains.
However, we find other interesting quantities, essentially regularized moments, that can be estimated and also provide useful information on entanglement of the two-photon state.
In the light of the possible applications of the HOM interference for time measurement~\cite{Giovannini857,Chen2019g, Scott2020}, we demonstrate that semiparametric methods offer an intriguing solution for model-independent estimation.

\section{Background}

\subsection{Basics of HOM interference}
The HOM effect consists in a two-photon interference occurring when these arrive at the same time on a beam splitter (BS) from separate ports~\cite{PhysRevLett.59.2044}.
This results in a suppression of the observed coincidence rate $C$, as measured by photon detectors at the two BS outputs.
The effect is generally studied by scanning the relative delay $\tau$ in the arrival times, producing an interference figure $C(\tau)$, modulated from $C(\tau)=C_0$ for long delays to a minimum achieved for $\tau=0$; its value is dictated by the reflectivity $R$ and trasmittivity $T=1-R$ of the BS, as well as the spectral properties of the photons.
For instance for a symmetric BS, with $R=1/2$ and spectrally indistinguishable photons, one would expect $C(0)=0$, thus the deviation from this condition is often adopted as a measurement of the level of indistinguishability~\cite{branczyk2017hongoumandel}.

More in detail, the coincidence profile $C(\tau)$ can be written as
\begin{equation}
\label{homdip}
    C(\tau) =C_0 \left( 1 - v \tilde f(\tau)\right),
\end{equation}
where $v=2RT/(R^2+T^2)$. We can write $\tilde f(\tau)=\int\, e^{i\omega\tau}f(\omega)d\omega$.
The function $f(\omega)$ is related to the spectral two-photon wavefunction (or joint spectral amplitude) $\Phi(\omega_1,\omega_2)$ as~\cite{PhysRevLett.115.193602}:
\begin{equation}
\label{fomega}
f(\omega)= \frac{1}{2} \int d\Omega\, \Phi^*\left(\frac{\Omega+\omega}{2},\frac{\Omega-\omega}{2}\right)\Phi\left(\frac{\Omega-\omega}{2},\frac{\Omega+\omega}{2}\right),  
\end{equation}
{i.e} $f(\omega)$ is the marginal symmetrised wavefunction along the direction $\omega=\omega_1-\omega_2$. Since this function is Hermitian $f(-\omega)=f^*(\omega)$ its Fourier transform $\tilde f(\tau)$ is guaranteed to be real. Concerning the phase of $f(w)$, this can not include quadratic terms, due to Hermicity, and linear terms can be accounted for simply by translating the origin of the delays $\tau$; since higher-order terms are usually small, we can consider $f(\omega)$ to be real, hence $\tilde f(\tau)$ is an even function.
Details are found in Appendix~\ref{app:A}.

\subsection{Semiparametric estimation}

The estimation of any parameter $\theta$ connected to $f(\omega)$ without assuming a specific form for it, implicitly relies on the knowledge of many other parameters $\boldsymbol{\eta} = [\eta_1, \eta_2,...,\eta_M]$ needed for describing the spectral function.
An example could be the estimation of one given moment of $f(\omega)$, with all the others acting as nuisance parameters.
Here the problem lies in the fact that $M$ could be too large for practical purposes, with the genuine semiparametric setting being achieved when $M\rightarrow \infty$.
In the standard parametric approach, it would be natural to try and write the Fisher information matrix $\mathbf{\cal F}$ of the vector of parameters $[\theta,\boldsymbol{\eta}]$ and obtain the Cram\'er-Rao bound (CRB) for an unbiased estimator of $\theta$ by means of its inversion:
\begin{equation}
\label{CRB}
    \Delta^2 \check{\theta} \geq \frac{1}{N} \left({\bf{\cal F}}^{-1}\right)_{11},
\end{equation}
with $N$ being the number of repetitions of the experiment.
When the dimension of the Fisher matrix $M$ is large, inversion could be difficult and prone to numerical instabilities, or even unfeasible in the semiparametric limit due to its large size.
This bound is written in terms of the classical Fisher information and thus pertains to a specific choice of the measurement, and should not be confused with the quantum version, which is independent on the setting of the experiment.

The theory of semiparametric estimation assists us in obtaining an expression for the bound \eqref{CRB} without manipulating large, formally infinite-dimensional, matrices. The evaluation of the CRB is based on geometrical considerations: instead of evaluating and inverting the Fisher information matrix of multiple parameters, the optimal bound is obtained by Hilbert space methods~\cite{bickel1998efficient,Tsiatis2006}.
While the complete details of the theory are quite technical, we rely on the treatment in~\cite[Sec.~II-III]{PhysRevResearch.1.033006}, which deals with the semiparametric estimation of the moments of an incoherent light source with an arbitrary spatial distribution. Here we adopt the same approach, but we have the time delay variable $\tau$ rather than a spatial distribution. 

Consider a generic parameter defined as $\theta=\int f(\omega) \vartheta(\omega)d\omega$, by means of a known function $\vartheta(\omega)$.
Our aim is to estimate $\theta$ relying only on its definition, but without assuming a particular functional form for $f(\omega)$.
Since in practice we have access to the function $\tilde f(\tau)$ in the conjugate domain of times, we can write:
\begin{equation}
\begin{aligned}
\label{semiparaest}
    \theta=&\int d\omega\, \vartheta (\omega) \frac{1}{2\pi} \int d\tau\,e^{-i\omega \tau}\tilde f(\tau)\\
    =& \int d\tau\,  \tilde f(\tau) \tilde \vartheta(\tau),
    \end{aligned}
\end{equation}
meaning that the parameter $\theta$ can equivalently be determined by the known function $\tilde \vartheta(\tau)$ in the time domain.
This parameter is, strictly speaking, defined by integrating in $\tau$ over the whole real line, but in practice we will approximate this with an integral in a finite range $[-T,T]$, symmetric around $0$.
In the following, for convenience, instead of $\theta$ we consider the parameter
\begin{equation}
\begin{aligned}
\label{eq:referee}
\theta' &= \int d\tau\, \tilde \vartheta(\tau)\frac{C(\tau)}{C_0} \\ 
&=-v\, \theta+\int d\tau\, \tilde \vartheta(\tau),
\end{aligned}
\end{equation}
which is more closely related to experimental data, i.e. the coincide profile $C(\tau)$.

Formally we can introduce a `detector space' $\mathcal{T} \subset \mathbb{R}$~\cite{PhysRevResearch.1.033006} which describes the possible settings of the detection, in our case, the time delay $\tau \in \mathcal{T}$.
We then introduce two measures on this space, \textit{i.e.} two ways of weighting the settings: $d\mu(\tau)$ which considers the actual experimental choices, and a random measure $dn(\tau)$, which accounts for the registered intensities, and presents Poisson statistics with mean $d\bar n(\tau)$
\footnote{Strictly speaking, these are already written in infinitesimal form, as they would appear inside an integral, but the measure is actually defined on any subset of the detector space.}.

While in optical imaging an array of spatially distributed photodetectors is often considered continuous~\cite{PhysRevResearch.1.033006}, in our HOM scenario the time delays $\tau$ at which measurements are performed are necessarily discrete and our analysis is carried out with a finite number of delays $\tau_i$, equally spaced at intervals of size $\delta \tau$.
We can then obtain insights on how to apply parametric methods by exploring the discrete setting first, and then understand how to recover the continuous limit.
In the discrete case $d\mu(\tau)$ is a comb of Dirac measures, $d\mu(\tau)=\sum_{i=1}^{i_{\mathrm{max}}}\delta(\tau-\tau_i) d\tau$.
This implies that for our experiment we can write the average Poisson measure as $d\bar n(\tau)=C(\tau)d\mu(\tau)$.

To embrace the discrete nature of the detection, instead of considering the original parameter, we introduce a new parameter, an approximation of $\theta'$ obtained by a discretization of the integral, i.e. $ \theta_{\mathrm{\delta \tau}}' = \sum_{i=1}^{i_{\mathrm{max}}} C(\tau_i)\tilde{\vartheta}_{\delta \tau}'(\tau_i)$, where we have defined $\tilde{\vartheta}'_{\delta \tau}(\tau)=\tilde{\vartheta}(\tau)\delta \tau/C_0$, such that this definition correctly reproduces the integral~\eqref{eq:referee} in the limit $\delta \tau \to 0 $.

A natural estimator for the discretized parameter $\theta'_{\delta \tau}$ is the following:
\begin{equation}
\begin{aligned}
\label{eq:spsum}
    \check{\theta}'_{\delta \tau} 
    = \int\,d n (\tau) \tilde \vartheta'_{\delta \tau}(\tau)
    = \sum_{i=1}^{i_{\mathrm{max}}} \check{C}(\tau_i)\tilde \vartheta'_{\delta \tau}(\tau_i),
   \end{aligned}
\end{equation}
where $\check{C}(\tau_i)$ is the observed number of coincidences at a given delay $\tau_i$, which is a minimum-variance and unbiased estimator for the true Poisson rate $C(\tau_i)$.
Crucially, the overall estimator $\check{\theta}'_{\delta \tau}$ assumes no a priori form of the profile $C(\tau)$.
The variance of this estimator is
\begin{equation}
\begin{aligned}
\label{eq:spvar}
    \Delta^2 \check{\theta}'_{\delta \tau} 
     &=\sum_{i=1}^{i_{\mathrm{max}}}C(\tau_i) \left(\tilde \vartheta'_{\delta \tau} (\tau)\right)^2
\end{aligned}
\end{equation}
and the variance on the estimator for the original (but discretized) parameter $\theta_{\delta \tau}$ is then obtained by error propagation via \eqref{eq:referee} as $\Delta^2 \check{\theta}_{\delta \tau} =\Delta^2\check{\theta}'_{\delta\tau}/v^2$.
Notice how this is dictated by $C_0$, rather than by the overall collected counts.
Eq.~\eqref{eq:spvar} simply amounts to the variance of a linear combination of independent Poisson random variables.

We can go back to the original parameter $\theta$ and consider the ideal limit of a continuous set of observations, i.e. a measure $d\mu(\tau)=d\tau$ on detector space.
To make a fair comparison and obtain the CRB in this limit, we need to fix the total resources and hence $dn(\tau)/dt$ must take the form of a coincidence density, which coincides with the flat distribution $C_0/\delta \tau$ far from the dip, thus we take $d\bar{n}(\tau)=\left( C(\tau)/\delta \tau \right) d\tau$.
If the full set of parameters $[ \theta, \boldsymbol{\eta} ]$ is large enough to parametrize arbitrary profiles $C(\tau)$ (i.e. if the tangent space of the statistical model is full-dimensional) we can derive the following semiparametric CRB~\cite{PhysRevResearch.1.033006}, in this continuous setting
\begin{equation}
\label{eq:truevar}
    \Delta^2\check{\theta'} \geq \int d\tau \frac{C(t)}{\delta \tau}  
    \left(\tilde \vartheta'_{\delta \tau} (\tau)\right)^2,
\end{equation}
where the right hand side is a continuous version of the variance~\eqref{eq:spvar} for the discretized parameter; see Appendix~\ref{app:semiparderiv} for more details.

We stress that the equations~\eqref{semiparaest} \emph{define} our parameter of interest $\theta$ in relation to $f(\omega)$ or $\tilde{f}(\tau)$, but clearly if one can obtain a (nonparametric) estimate of the function $\tilde{f}(\tau)$ then \eqref{semiparaest} can be used to estimate $\theta$ without relying on a parametric model for $\tilde{f}(\tau)$.
This estimation strategy has the advantage of reducing possible biases due to discretization.
More concretely, we first obtain an estimate $\check{C}(\tau)$ of the continuous density by interpolation of the observed coincidences $\check{C}(\tau_i)$ and then the estimator $\check{\theta}'$ is obtained as
\begin{equation}
    \label{eq:truesp}
    \check{\theta}' =\int d\tau \frac{ \check{C}(t)}{\delta \tau}  
    \tilde \vartheta'_{\delta \tau} (\tau).\\
\end{equation}

\subsection{Hermite-Gauss parameter estimation} 
For a straightforward characterisation of $f(\omega)$, it would be convenient to extract 
its moments, defined as the integrals of $\omega^nf(\omega)$
over the frequency space. In this case, the associated semiparametric estimators are the $n$-th derivative of the Dirac delta function, according to \eqref{semiparaest}. Since the corresponding experimental estimate would not be a well-defined quantity, the semiparametric approach cannot be applied in similar instances.
Instead, we can focus on the decomposition of $f(\omega)$ in Hermite-Gauss (HG) functions:
\begin{equation}
\label{hgfunctions}
    \HG_n(\omega) = e^{-\xi^2\omega^2}H_n(\xi\omega),
\end{equation}
where $H_n(x)$ is the $n$-th Hermite polynomial and we have $\tilde{\HG}_{n}(\tau) = (-\I)^n e^{-\frac{\tau^2}{4 \xi^2}} \frac{\tau^n}{\xi^{n+1}}$ \footnote{Notice that our definition of the HG functions differs from the usual eigenfunctions of the Fourier transform, as we have used the multiplying factor $e^{-\xi^2\omega^2}$ instead of $e^{-\xi^2\omega^2/2}$.} and $\xi$ is a positive scale parameter with units of time. These quantities measure the contribution of modulating terms in the spectral function \eqref{fomega}, with an exponential providing the necessary regularity.
Due to the Hermitian symmetry imposed by \eqref{fomega}, we expect all odd-$n$ terms to vanish.
The extraction of any particular HG component, independently on the others by making no assumptions on the density profile $C(\tau)$, is a semiparametric problem. In particular, we can exploit the fact that, for a separable wavefunction $\Phi(\omega_1,\omega_2)=\phi_1(\omega_1)\phi_2(\omega_2)$ the following condition must hold:
\begin{equation}
\label{mathias}
    h_{2n} =(-1)^n\int \HG_{2n}(\omega)f(\omega)\geq 0.
\end{equation}
This can be demonstrated as follows: we consider a factorable state $\Phi(\omega_1,\omega_2)=\phi_1(\omega_1)\phi_2(\omega_2)$, thus we can write $f(\omega)={\int S(\omega+\Omega)^*S(\omega-\Omega)d\Omega}$, with $S(2x)=\phi_1(x)\phi_2(x)^*$. By Khintchine's criterion~\cite{Ushakov2011}, the function $f(\omega)/f(0)$ is a characteristic function of a probability distribution. For all characteristic functions $\chi(\omega)$, Mathias' theorem~\cite{Mathias1923} has that $(-1)^n\int \chi(\omega)HG_{2n}(\xi\omega)\geq 0$ for all integers $n$, and for all $\xi>0$. Since $f(0)>0$, our separability condition~\eqref{mathias} holds.
The result remains valid also in the more generic case of a mixed state, for which $f(\omega)$ is a convex combination of integrals in the same form, which is still a characteristic function. 

The quantities $h_{2n}$ constitute the parameters we estimate by means of the semiparametric approach discussed in the previous section: in the following the previous abstract parameter $\theta$ will correspond to $h_{2n}$ for different $n$.
The semiparametric estimation of one parameter $h_{2n}$ violating the positivity condition acts as witness of spectral entanglement.
We notice that this is equivalent to observe values $\tilde f(\tau)<0$, {i.e} $C(\tau)>C_0$, a well-known fact~\cite{Wang2006a,Fedrizzi_2009}.
However, our approach recasts this entanglement criterion in different quantitative terms, remarkably, by considering the whole shape of the interferogram, rather than individual points.
We thus focus our attention of one such parameters $h_{2n}$, without assuming any parametric model for the coincidence profile $C(\tau)$.

\begin{figure}[ht!]
    \includegraphics[width=\columnwidth]{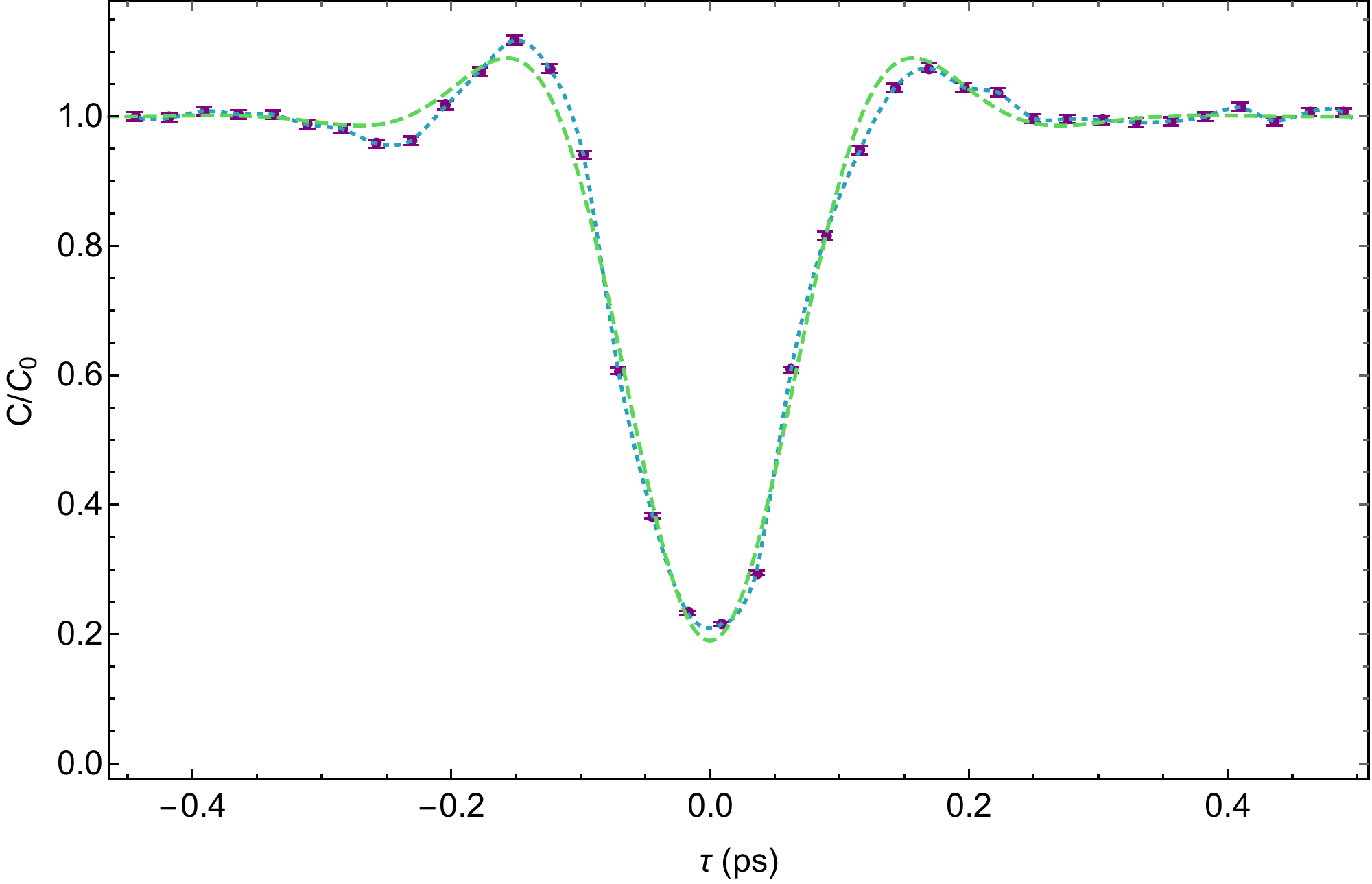}
    \caption{Coincidence dip profile, normalized to $C_0$ coincidences, obtained scanning the relative delay between the two photons when arriving at the BS. The photon pairs are detected through avalanche photodiodes after passing two interference filters (fourth order superGaussian profile, $7.3$ nm width). The coincidences counts are collected in $5$ s, with $C_0=4653$ coinc.
    The dotted line is the interpolation, the dashed line is a fit with the function reported in Appendix~\ref{app:A}.}
    \label{fig:dip}
\end{figure}

\section{Results}
Our set up is the standard HOM interferometer in which two photons from a downconversion crystal ($\beta$ barium borate, 3 mm length, degenerate type-I phase matching at $\lambda$=810 nm) arrive on a beam splitter; this was chosen with reflectivity $R\sim2/3$, thus setting the visibility in \eqref{homdip} to $v=0.81$.
The use of a CW pump makes the wavefunction almost monocromatic along $\Omega$, as enforced by energy conservation, while two interference filters define the wavefunction in the $\omega$ direction, since the intrinsic bandwidth of the downconversion emission, as dictated by the crystal length, is much wider.
The HOM dip shape $C(\tau)/C_0$ has been reconstructed at different points, as shown in Fig.~\ref{fig:dip}. The delay $\tau$ was controlled by means of a translation stage.
The interference figure is collected at a sampling rate of $\delta\tau=13.4$ fs, as reported in Fig.\ref{fig:dip}, and then interpolating by means of third-order polynomials.
This constitutes the data set we use for estimation of generalised momenta of order $n=0, 2, 4$, seconding the expected symmetry. 

\begin{figure}[h!]
    \includegraphics[width=\columnwidth]{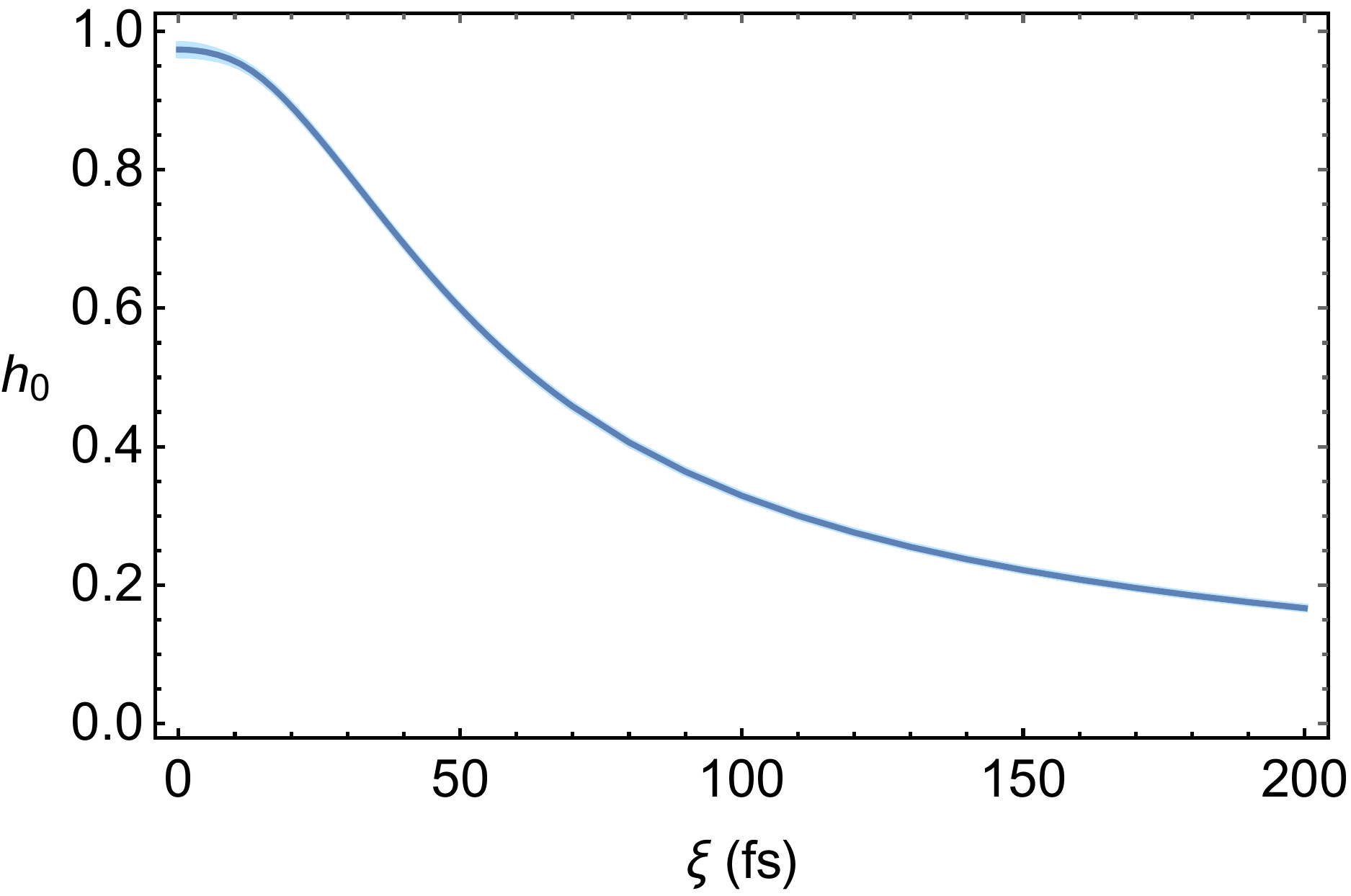}
        \includegraphics[width=\columnwidth]{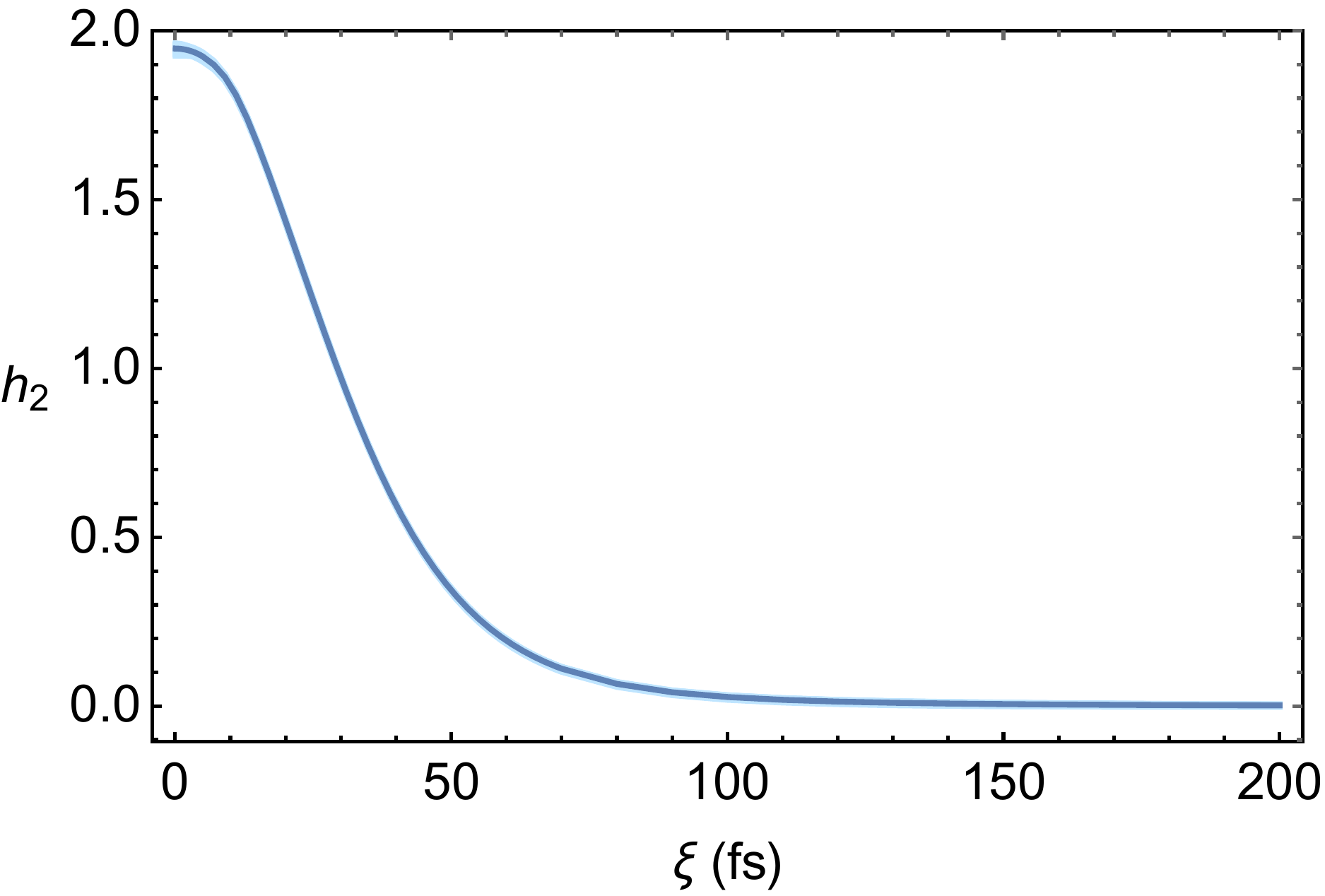}
    \includegraphics[width=\columnwidth]{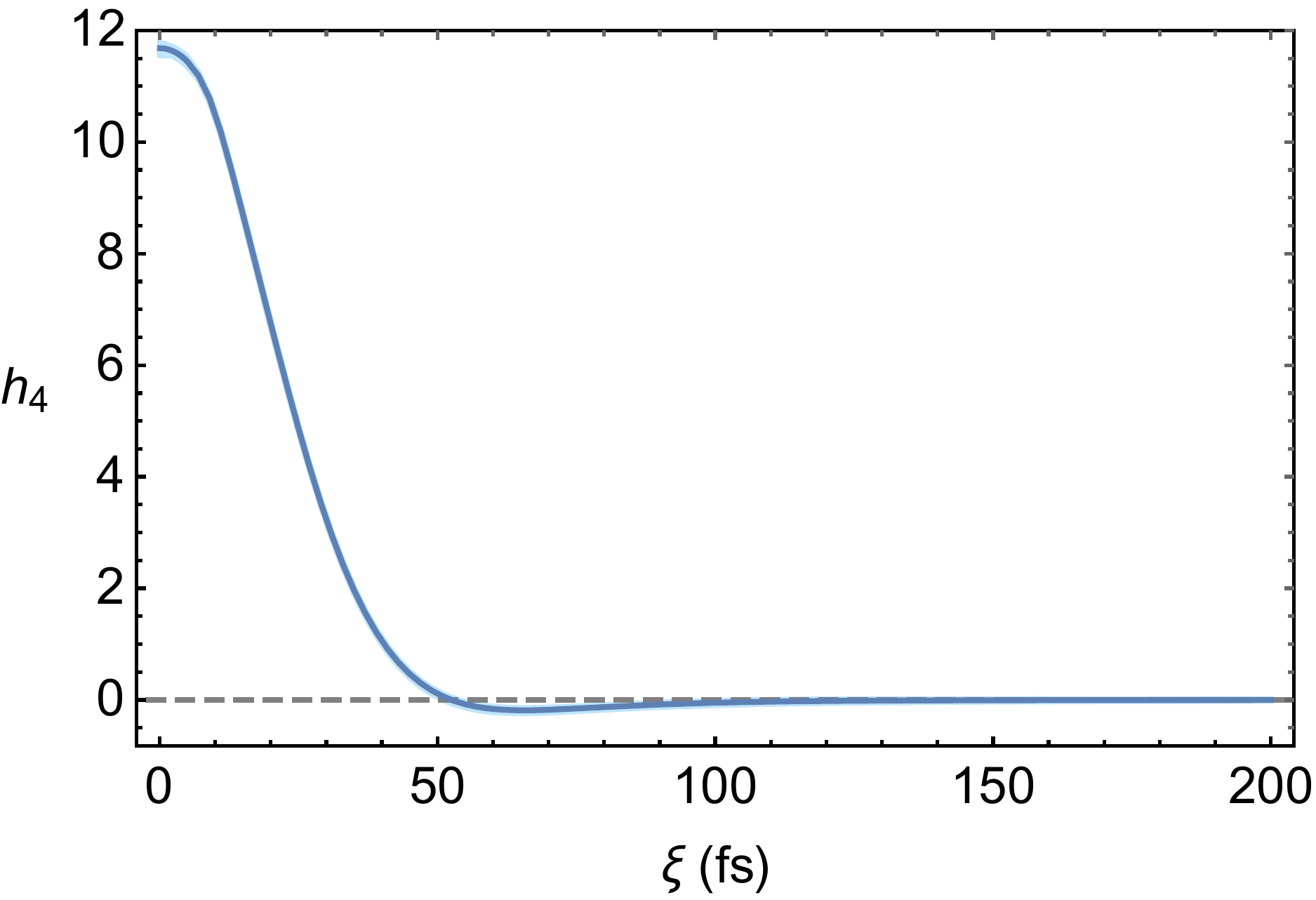}
    \caption{Estimation of HG parameters obtained with the semiparametric method from the measured data as function of the width of the HG functions determined by $\xi$.The plots show the estimate for a) the zero-order one, b) the second-order term and in c) the fourth-order one which shows a negative region witnessing the presence of entanglement. The shading under the curve indicates the uncertainty on the estimated parameters.}
    \label{fig:mean}
\end{figure}

In Fig.~\ref{fig:mean} we plot the semiparametric estimates of $h_{n}$, as a function of the parameter $w$, as obtained by the integral estimator in \eqref{eq:truesp} based the interpolated function: the semiparametric method offers reliable estimates, and $h_4$ is the first to witness the presence of entanglement in the state, taking negative values for a wide range of $\xi$.

\begin{figure}[h!]
    \includegraphics[width=\columnwidth]{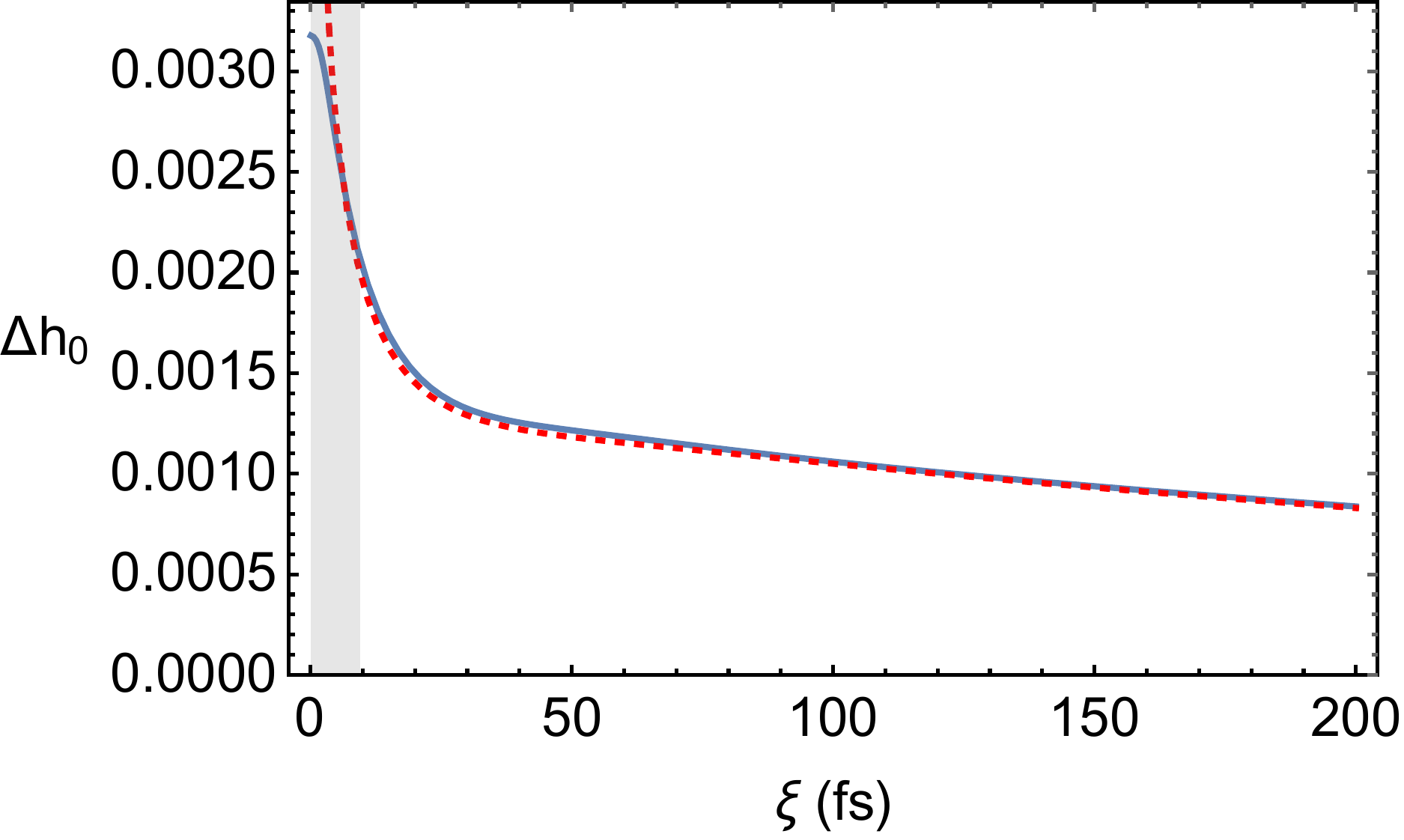}
        \includegraphics[width=\columnwidth]{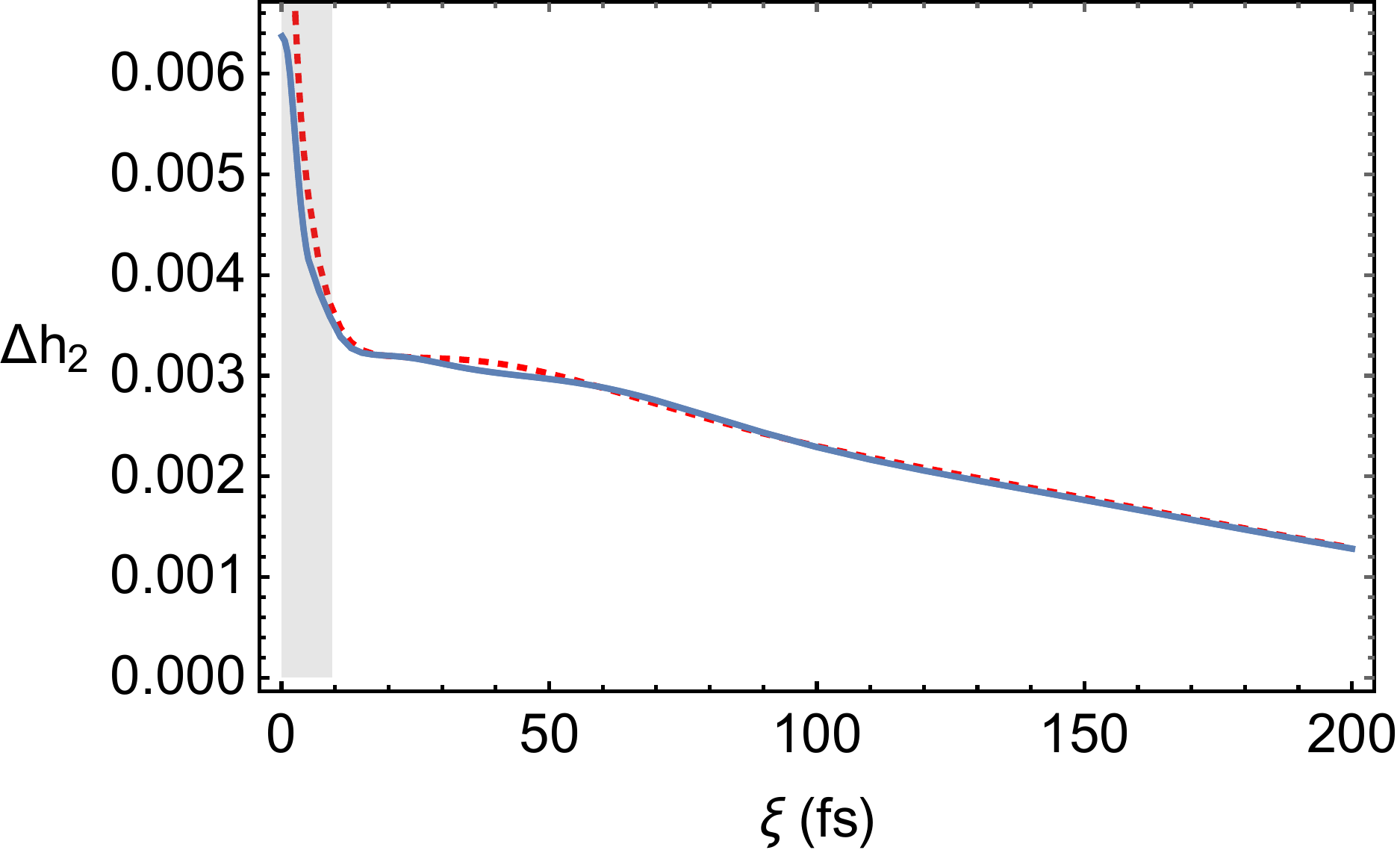}
    \includegraphics[width=\columnwidth]{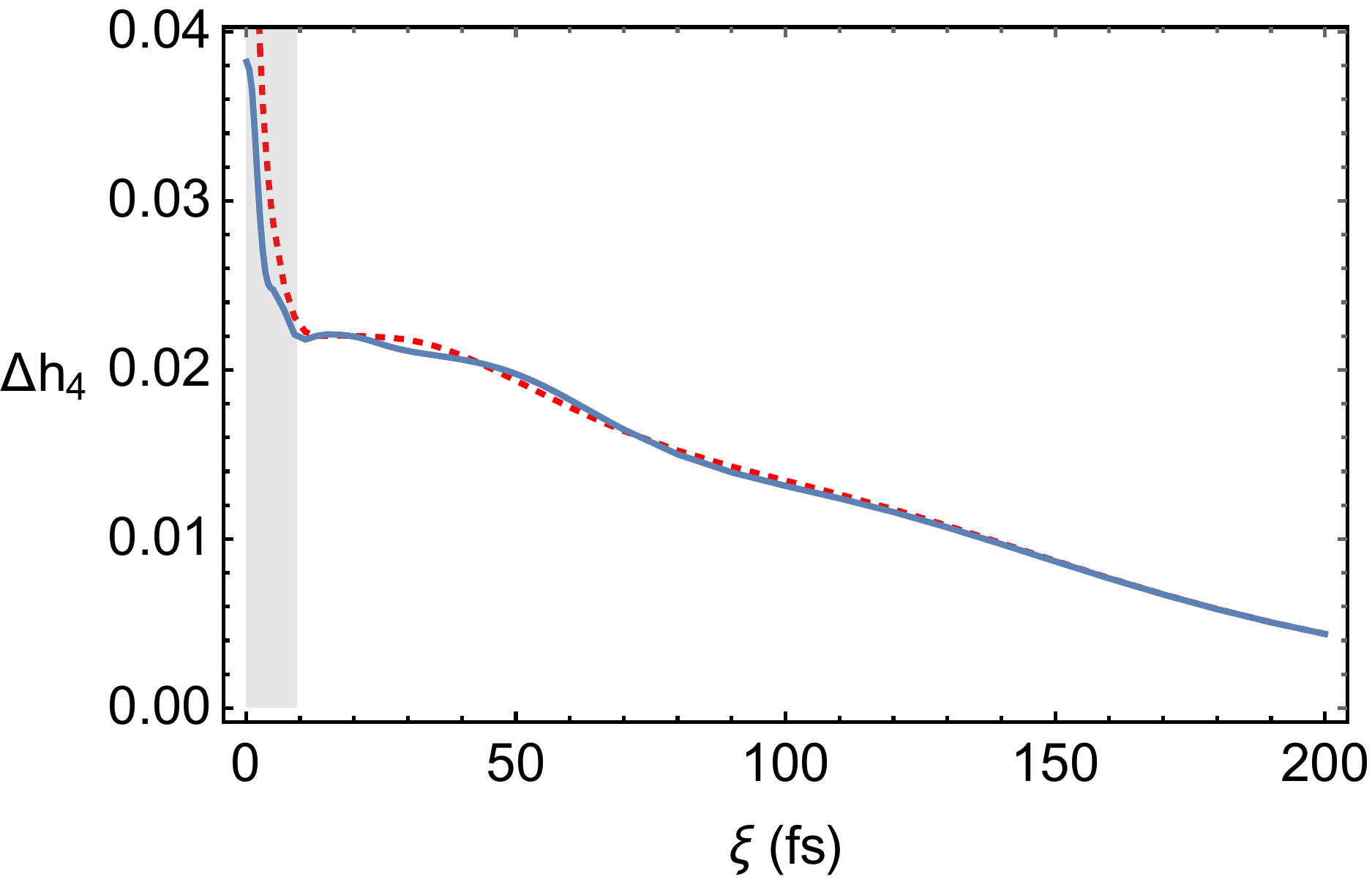}
    \caption{Uncertainty on the estimation of HG parameters obtained with a Monte Carlo simulation over $1000$ repetitions considering Poissonian noise to the measured data. The red dotted line shows the relative CRB, Eq.\eqref{eq:truevar}. The shaded grey area identifies the one in which a bias can be expected, and may show as a violation of the CRB }
    \label{fig:variance}
\end{figure}

The corresponding uncertainty are analysed in Fig.~\ref{fig:variance}, which show the standard deviation $\Delta h_n$ on the estimated parameters. This is assessed by means of a bootstrap method, consisting in Monte Carlo repetitions of the experiment, based on the registered experimental counts; this smoothly accounts for the contribution to the uncertainty due to the interpolation step.
This reveals that, despite the curves of the average values appear regular, a bias occurs, manifesting as a violation of the semiparametric CRB in \eqref{eq:truevar}.
We hence conclude that our estimator \eqref{eq:truesp} is not an unbiased estimator of the parameter, since the \emph{discrete} nature of the original data still affects it in the interpolation needed to obtain the estimator $\check{C}(\tau)$ for the continuous density. 
This means that if we keep spending resources to increase the precision on the punctual estimates of the rates $C(\tau_i)$ without reducing $\delta \tau$, the approximated parameter will eventually reveal its difference from the true one.
Our statistical model assumes that uncertainties on the data is purely statistical, whereas the interpolation is affected by errors of a different kind.
This idea that one needs to balance between interpolation error and punctual statistical errors is quite and general, and was recently reported in the context of function estimation with multiple phase measurements~\cite{Gianani2020}. 

The discrepancy with the CRB depends strongly on the value of $\xi$. We can derive an argument illustrating what is the region in which we can neglect our bias. 
In fact, the width of $\HG_{2n}(\omega)$ increases as $\xi$ is reduced, and the spacing $\delta \tau$ eventually becomes too large to capture variations---this effectively imposes an low-pass filter.
Since these functions have no compact support, we cannot use a Shannon-Nyquist criterion rigorously to evaluate the quality of the sampling; however, we can take as a guiding principle the fact that the sampling frequency $1/\delta \tau$ must exceed the width of the Gaussian in~\eqref{hgfunctions}, leading to $\xi>\delta \tau/\sqrt{2}$. Below this value, we can not rely on our estimate, even if an interpolation is used.
Conversely, for a target parameter, the bandwidth of the corresponding function determines what sampling step can be judged satisfactory.

The interval in which $h_4$ witness entanglement, on the other hand, is safely outside this unreliability region: we can then assess the significance of the violation of the condition $h_4\geq0$ by considering the signal-to-noise ratio $R_4=h_4/\Delta h_4$. This reveals the validity of our witness.

\begin{figure}[h!]
    \includegraphics[width=\columnwidth]{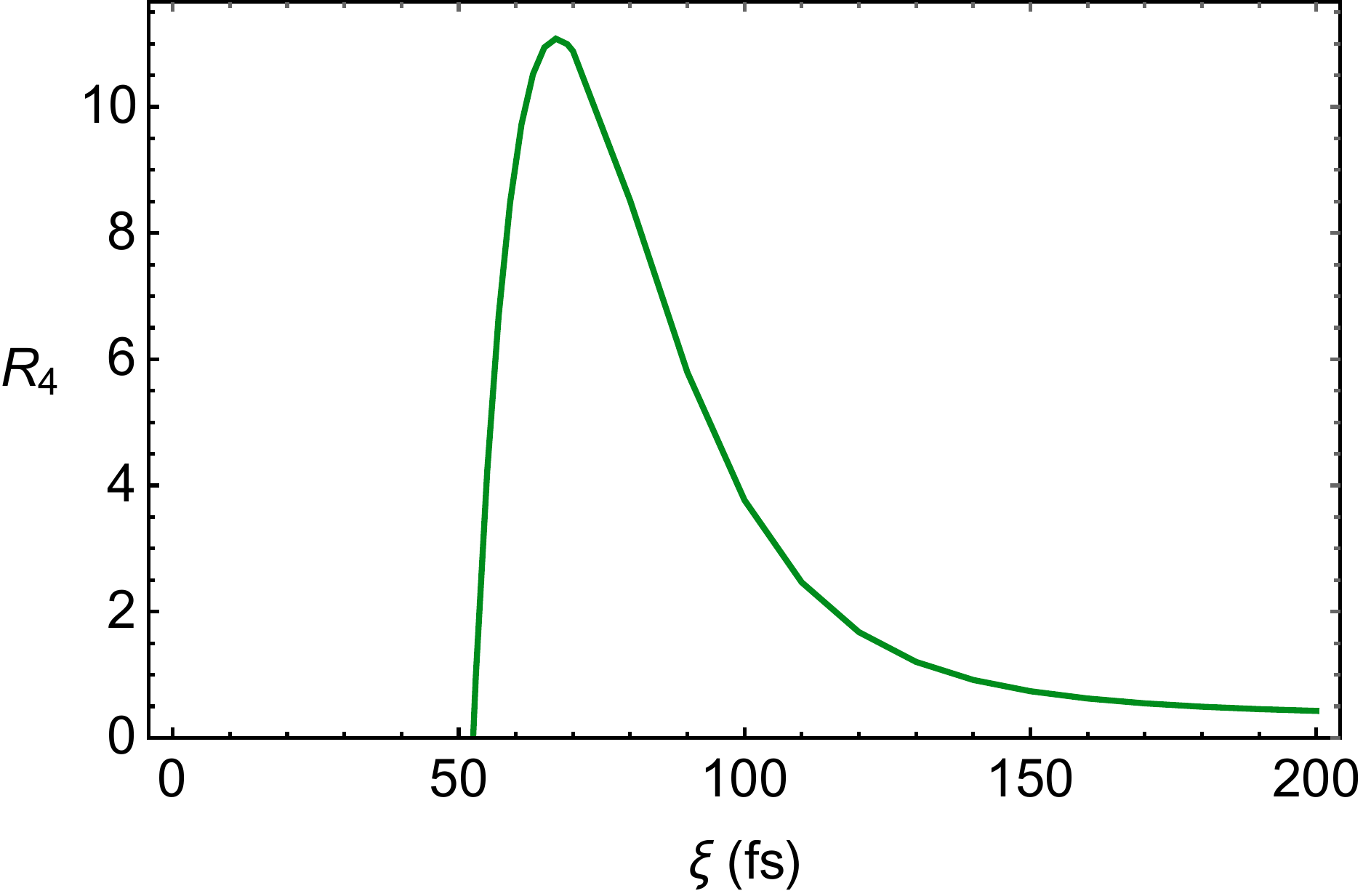}
    \caption{Negative signal to noise ratio $R_4$ relative to the fourth-order HG parameter $h_4$, indicating the significance of the negative region in Fig.~\ref{fig:mean}c.}
    \label{fig:sensitivity}
\end{figure}

\begin{figure}[h!]
    \includegraphics[width=\columnwidth]{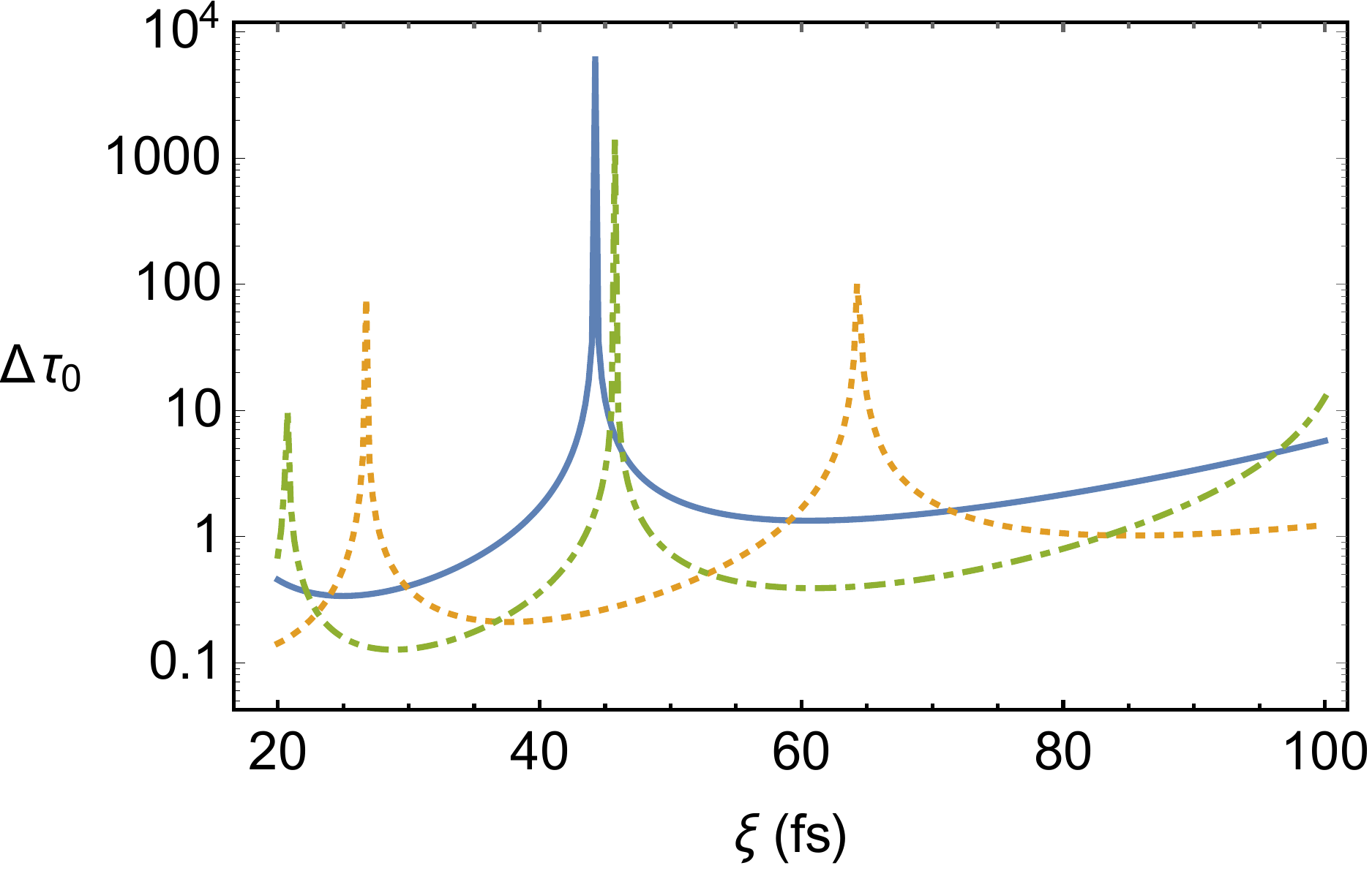}
    \caption{Uncertainty on the estimation of an extra delay $\tau_0$ obtained through $h_0$ (blue solid line), $h_2$ (orange dotted line) and $h_4$ (green dashed-dotted line) as a function of $\xi$.}
    \label{fig:time}
\end{figure}

The estimates of the parameters \eqref{hgfunctions} can serve the purpose of using the HOM profile for the measurement of small time delays, without resorting to fitting the coincidence curve to a specific model~\cite{Giovannini857}. Considering an extra delay $\tau_0$ shifting the coincidence curve as $\tilde f(\tau-\tau_0)$, we can evaluate what uncertainty $\Delta\tau_0$ can be obtained, by measuring the three Hermite-Gauss parameters $h_0$, $h_2$, or $h_4$. For small shifts, these can be evaluated as $\Delta\tau_0 =\left\vert  \Delta h_i/\left.\partial_{\tau_0}h_i\right\vert_{\tau_0=0} \right\vert$, which are shown in Fig.\ref{fig:time} as a function of $\xi$. As a general rule, higher modes provide lower uncertainties, when $w$ is properly set; notice that this optimisation for $h_0$ is akin to a standard fit enforcing a Gaussian shape. It should be noted that higher HG terms would be less reliable: since they consider modulations with shorter periods in $\tau$, they would be more affected by fluctuations of the level of the signal.
Further, since the bandwidth of the HG functions grows with the order, finer sampling could be required.
Our analysis shows that, with the collected number of events, one can reduce the statistical uncertainty to the point where instrumental effects---notably, the reproducibility of the translation stage movements---become the main source of error. On the other hand, this technique requires a complete scan of the coincidence profile, as well as a calibration step at $\tau_0=0$, differently from model-dependent techniques~\cite{Lyons2018}.     

\section{Conclusions}

We have presented a semiparametric analysis of the Hong-Ou-Mandel interference profile. The use of Fourier transforms curtails the adoption of semiparametric methods, nevertheless an analysis in terms of Hermite-Gauss functions can be effectively carried out.
This brings about a reinterpretation of a standard entanglement witness in terms of spectral properties.
The possible use of this analysis for delay measurements has been illustrated.

In this respect, tools from statistical classical and quantum estimation theory have already proven to be extremely useful for analysing and engineering metrological schemes based on HOM interference~\cite{Lyons2018,Chen2019g,Harnchaiwat2020,Scott2020,Scott2021,Ndagano2021}.
Accordingly, we expect that delving into more advanced statistical methods will also bring further insights to such applications.

\section{Acknowledgments} 

We acknowledge useful discussion with F. Sciarrino.
V.C., I.G., and M.B. acknowledge support from the FET-OPEN-RIA grant STORMYTUNE (Grant Agreement No. 899587) of the European Commission. V.C. is supported by the Amaldi Research Center funded by the Ministero dell’Istruzione dell’Università e della Ricerca (Ministry of Education, University and Research) program “Dipartimento di Eccellenza” (CUP:B81I18001170001).
F.A. acknowledges financial support from the National Science Center (Poland) Grant No. 2016/22/E/ST2/00559.

\appendix
\section{Quantum mechanical model}\label{app:A}

A generic pure state of two spectrally entangled photons is
\begin{equation}
\ket{\Phi} = \int d\omega_1 d\omega_2 \Phi(\omega_1,\omega_2) a_1^\dag(\omega_1) a_2^\dag(\omega_2) \ket{0},
\end{equation}
where $\Phi(\omega_1,\omega_2)$ is the joint spectral amplitude and the integrals extend from $-\infty$ to $\infty$, since we assume that $\Phi(\omega_1,\omega_2)$ is strongly peaked.

The probability of a coincidence for this state, assuming $v=1$ for simplicity, is obtained as~\cite[Eq.~(40)]{branczyk2017hongoumandel}
\begin{multline}
\label{eq:probCoinc}
p_C(\tau)=\frac{1}{2} \biggl[ 1 - \frac{1}{2} \int d \omega e^{-\I \omega \tau} \int d \Omega \\
\Phi^* \left(\frac{\Omega+\omega}{2},\frac{\Omega-\omega}{2} \right) \Phi\left(\frac{\Omega-\omega}{2},\frac{\Omega+\omega}{2} \right)  \biggr].
\end{multline}

For biphoton states generated by spontaneous parametric downconversion (SPDC), the joint spectral amplitude can be expressed as: 
\begin{equation}
\Phi(\omega_1,\omega_2)= \alpha( \omega_1+\omega_2 ) \phi(\omega_1,\omega_2)
\end{equation}
where $\phi(\omega_1,\omega_2)$ is the phase-matching function and $\alpha(\omega_1+\omega_2)$ is the pump spectrum.
For CW pumping we have that the frequencies of the two photons are perfectly anticorrelated, i.e. $\alpha(\omega_1,\omega_2) \propto \delta(\omega_1 + \omega_2 - 2 \bar{\omega})$, where $2 \bar{\omega}$ is the pump frequency.
In our experiment, the phase matching function is wide along the $\omega$ axis, thus it will be governed by the interference filters we have used to select the modes. We have a factorized super-Gaussian phase-matching function~\cite{sbroscia}:
\begin{equation}
\phi(\omega_1,\omega_2) \propto e^{-\frac{(\omega_1-\bar\omega)^4}{2 \sigma_0^4}} e^{-\frac{(\omega_2-\bar\omega)^4}{2 \sigma_0^4}},
\end{equation}
where everything is normalized such that $\int d\omega_1 d\omega_2 |\Phi(\omega_1,\omega_2)|^2 = 1$. 
While the analytical expressions for the function $\tilde{f}(\tau)$, and thus $C(\tau)$ is hard to obtain, we found that it is well approximated as~\cite{flamini} $\tilde{f}(\tau)=e^{-\frac{\tau^2}{2\sigma^2}} \operatorname{sinc} \left( \pi B \frac{\tau}{\sigma} \right)$, where the two parameters $\sigma$ and $B$ can be fitted from experimental data; this has also been employed to locate the zero position of the translation stage. By taking the inverse Fourier transform of this fitted function we can also find an expression for the function $f(\omega)$.

Finally, it is important to notice that we have written~\eqref{eq:probCoinc} as the \emph{probability} of a coincidence event, but this assumes a deterministic generation of the state $\ket{\Phi}$.
In reality, SPDC is a probabilistic process and the number of recorded coincidences in a fixed amount of time $T$, for a given delay $\tau$, approximately follows a Poisson distribution with average $ N_0 p_C(\tau)$, where $N_0$ is the generation rate. Thus, we can treat the data at each point $\tau_i$ as an independent Poisson variable.

\section{Semiparametric bound for arbitrary coincidence profile}
\label{app:semiparderiv}

We consider the segment on the real line $\tau \in [-T,T]$ with the standard Lebesgue measure as our detector space.
As explained in~\cite{PhysRevResearch.1.033006} one can introduce a Hilbert for detector-space functions to evaluate the semiparametric CRB.
We first introduce the scalar product
\begin{equation}
\langle h_1 , h_2 \rangle = \int_{-T}^T h_1(\tau) h_2(\tau)  d\bar{n}(\tau),
\end{equation}
where $d \bar{n}(\tau) = \frac{C(\tau)}{\delta \tau} dt$ represents the infinitesimal intensity measure, i.e. $\frac{C(\tau)}{\delta \tau}$ is the non-normalized density.
Here $C(\tau)$ is the true (but arbitrary) coincidence profile and the factor $\delta \tau$ is needed to fix the same amount of resources as in the discrete model.

The Hilbert space is then defined as the set of square summable functions 
\begin{equation}
    \mathcal{H} = \left\{ h(\tau) : \langle h , h \rangle < \infty  \right\}.
\end{equation}
For any parameter characterizing the coincidence profile $C(\tau)$ we can introduce the score function in detector space 
\begin{equation}
    S_{\theta_j} (\tau ) = \frac{ \partial }{ \partial \theta_j} \log C(\tau)
\end{equation}
which is the logarithmic derivative of the distribution (since $\delta \tau$ is parameter independent).
An important subspace of $\mathcal{H}$ is the tangent space $\mathcal{T}$, defined as the closure of $\mathrm{span}\{ S_j \}$, where $j$ can run over an uncountable set. 

Since we focus on a parameter that is a linear functional of the coincidence density $\int d\tau\,  \tilde C(\tau) \tilde \vartheta(\tau)$ (up to additive and multiplicative constants), the function $\tilde \vartheta(\tau) \in \mathcal{H}$ represents what is called an influence function. 
In general, the semiparametric CRB is obtained as $\langle \tilde \vartheta_{\mathrm{eff}} , \tilde \vartheta_{\mathrm{eff}} \rangle$, where $\vartheta_{\mathrm{eff}}$ is the so-called efficient influence, obtained by projecting any valid influence function onto the tangent space $\mathcal{T}$.
However, when the tangent space is so large that $\mathcal{T}=\mathcal{H}$ the projection is trivial, and the CRB is easily computed directly from the influence function.

The proof that the tangent space of arbitrary probability distributions is full-dimensional is found in~\cite[Example 1, Sec.~3.2]{bickel1998efficient}.
In our case we have not an arbitrary probability distribution, since the statistics is assumed to be Poissonian, but we have an arbitrary intensity density.
Following~\cite{PhysRevResearch.1.033006} we can use essentially the same method as in the standard case, using the Hilbert space of functions in the detector space instead of the standard statistical one.
The main difference is that the intensity distribution is not normalized, unlike a probability distribution.

In order to show that the family of arbitrary intensity measures has a tangent space that is maximally large, we rely on the concept of parametric submodels.
The idea underneath, setting its technicalities aside, is that we can introduce parametric models, such as 
\begin{equation}
    C_{\kappa}(\tau) = (1 + \tanh[ \kappa h(\tau) ] ) C(\tau),
\end{equation}
where the original true density is obtained at the value $\kappa=0$ and the score function is $S(\tau)=h(\tau)$.
Then, the set $\{S_j\}$ for the full model is the union of the tangent sets of all parametric submodels obtained for each $h\in \mathcal{H}$.
Intuitively, since we the functions $h(\tau)$ can be arbitrary this means that $\mathcal{T}=\mathcal{H}$. 

We should remark that, while we have assumed an arbitrary $C(\tau)$, this actually originates from the quantum mechanical model in~\eqref{homdip}, derived from the two-photon wave function $\Phi(\omega_1,\omega_2)$, or more generally a density operator $\rho( \omega_1,\omega_2, \omega'_1,\omega'_2)$.
In principle, this may impose constraints on $C(\tau)$ that could entail $\mathcal{T} \subset \mathcal{H}$: the semiparametric bound we have derived could be overestimating the true one. This, however, should not be expected to affect our claims on the presence of bias, as this is strongly connected to the data sampling rate, as revealed by downsampling the recorded coincidence profile. Elucidating the nature and implications of such constraints is an intriguing subject, left for future investigations.

\bibliography{biblio}

\providecommand{\noopsort}[1]{}\providecommand{\singleletter}[1]{#1}%
\begin{thebibliography}{58}%
\makeatletter
\providecommand \@ifxundefined [1]{%
 \@ifx{#1\undefined}
}%
\providecommand \@ifnum [1]{%
 \ifnum #1\expandafter \@firstoftwo
 \else \expandafter \@secondoftwo
 \fi
}%
\providecommand \@ifx [1]{%
 \ifx #1\expandafter \@firstoftwo
 \else \expandafter \@secondoftwo
 \fi
}%
\providecommand \natexlab [1]{#1}%
\providecommand \enquote  [1]{``#1''}%
\providecommand \bibnamefont  [1]{#1}%
\providecommand \bibfnamefont [1]{#1}%
\providecommand \citenamefont [1]{#1}%
\providecommand \href@noop [0]{\@secondoftwo}%
\providecommand \href [0]{\begingroup \@sanitize@url \@href}%
\providecommand \@href[1]{\@@startlink{#1}\@@href}%
\providecommand \@@href[1]{\endgroup#1\@@endlink}%
\providecommand \@sanitize@url [0]{\catcode `\\12\catcode `\$12\catcode
  `\&12\catcode `\#12\catcode `\^12\catcode `\_12\catcode `\%12\relax}%
\providecommand \@@startlink[1]{}%
\providecommand \@@endlink[0]{}%
\providecommand \url  [0]{\begingroup\@sanitize@url \@url }%
\providecommand \@url [1]{\endgroup\@href {#1}{\urlprefix }}%
\providecommand \urlprefix  [0]{URL }%
\providecommand \Eprint [0]{\href }%
\providecommand \doibase [0]{https://doi.org/}%
\providecommand \selectlanguage [0]{\@gobble}%
\providecommand \bibinfo  [0]{\@secondoftwo}%
\providecommand \bibfield  [0]{\@secondoftwo}%
\providecommand \translation [1]{[#1]}%
\providecommand \BibitemOpen [0]{}%
\providecommand \bibitemStop [0]{}%
\providecommand \bibitemNoStop [0]{.\EOS\space}%
\providecommand \EOS [0]{\spacefactor3000\relax}%
\providecommand \BibitemShut  [1]{\csname bibitem#1\endcsname}%
\let\auto@bib@innerbib\@empty
\bibitem [{\citenamefont {Hong}\ \emph {et~al.}(1987)\citenamefont {Hong},
  \citenamefont {Ou},\ and\ \citenamefont {Mandel}}]{PhysRevLett.59.2044}%
  \BibitemOpen
  \bibfield  {author} {\bibinfo {author} {\bibfnamefont {C.~K.}\ \bibnamefont
  {Hong}}, \bibinfo {author} {\bibfnamefont {Z.~Y.}\ \bibnamefont {Ou}},\ and\
  \bibinfo {author} {\bibfnamefont {L.}~\bibnamefont {Mandel}},\ }\bibfield
  {title} {\bibinfo {title} {Measurement of subpicosecond time intervals
  between two photons by interference},\ }\href
  {https://doi.org/10.1103/PhysRevLett.59.2044} {\bibfield  {journal} {\bibinfo
   {journal} {Phys. Rev. Lett.}\ }\textbf {\bibinfo {volume} {59}},\ \bibinfo
  {pages} {2044} (\bibinfo {year} {1987})}\BibitemShut {NoStop}%
\bibitem [{\citenamefont {Bra{\'n}czyk}(2017)}]{branczyk2017hongoumandel}%
  \BibitemOpen
  \bibfield  {author} {\bibinfo {author} {\bibfnamefont {A.~M.}\ \bibnamefont
  {Bra{\'n}czyk}},\ }\href@noop {} {\bibinfo {title} {{Hong-Ou-Mandel}
  interference}} (\bibinfo {year} {2017}),\ \Eprint
  {https://arxiv.org/abs/1711.00080} {arXiv:1711.00080 [quant-ph]} \BibitemShut
  {NoStop}%
\bibitem [{\citenamefont {Bouchard}\ \emph {et~al.}(2021)\citenamefont
  {Bouchard}, \citenamefont {Sit}, \citenamefont {Zhang}, \citenamefont
  {Fickler}, \citenamefont {Miatto}, \citenamefont {Yao}, \citenamefont
  {Sciarrino},\ and\ \citenamefont {Karimi}}]{Bouchard_2021}%
  \BibitemOpen
  \bibfield  {author} {\bibinfo {author} {\bibfnamefont {F.}~\bibnamefont
  {Bouchard}}, \bibinfo {author} {\bibfnamefont {A.}~\bibnamefont {Sit}},
  \bibinfo {author} {\bibfnamefont {Y.}~\bibnamefont {Zhang}}, \bibinfo
  {author} {\bibfnamefont {R.}~\bibnamefont {Fickler}}, \bibinfo {author}
  {\bibfnamefont {F.~M.}\ \bibnamefont {Miatto}}, \bibinfo {author}
  {\bibfnamefont {Y.}~\bibnamefont {Yao}}, \bibinfo {author} {\bibfnamefont
  {F.}~\bibnamefont {Sciarrino}},\ and\ \bibinfo {author} {\bibfnamefont
  {E.}~\bibnamefont {Karimi}},\ }\bibfield  {title} {\bibinfo {title}
  {Two-photon interference: the hong{\textendash}ou{\textendash}mandel
  effect},\ }\href {https://doi.org/10.1088/1361-6633/abcd7a} {\bibfield
  {journal} {\bibinfo  {journal} {Rep. Prog. Phys}\ }\textbf {\bibinfo {volume}
  {84}},\ \bibinfo {pages} {012402} (\bibinfo {year} {2021})}\BibitemShut
  {NoStop}%
\bibitem [{\citenamefont {Knill}\ \emph {et~al.}(2001)\citenamefont {Knill},
  \citenamefont {Laflamme},\ and\ \citenamefont {Milburn}}]{klm}%
  \BibitemOpen
  \bibfield  {author} {\bibinfo {author} {\bibfnamefont {E.}~\bibnamefont
  {Knill}}, \bibinfo {author} {\bibfnamefont {R.}~\bibnamefont {Laflamme}},\
  and\ \bibinfo {author} {\bibfnamefont {G.~J.}\ \bibnamefont {Milburn}},\
  }\bibfield  {title} {\bibinfo {title} {A scheme for efficient quantum
  computation with linear optics},\ }\href {https://doi.org/10.1038/35051009}
  {\bibfield  {journal} {\bibinfo  {journal} {Nature}\ }\textbf {\bibinfo
  {volume} {409}},\ \bibinfo {pages} {46} (\bibinfo {year} {2001})}\BibitemShut
  {NoStop}%
\bibitem [{\citenamefont {O'Brien}\ \emph {et~al.}(2003)\citenamefont
  {O'Brien}, \citenamefont {Pryde}, \citenamefont {White}, \citenamefont
  {Ralph},\ and\ \citenamefont {Branning}}]{cnotexperiment}%
  \BibitemOpen
  \bibfield  {author} {\bibinfo {author} {\bibfnamefont {J.~L.}\ \bibnamefont
  {O'Brien}}, \bibinfo {author} {\bibfnamefont {G.~J.}\ \bibnamefont {Pryde}},
  \bibinfo {author} {\bibfnamefont {A.~G.}\ \bibnamefont {White}}, \bibinfo
  {author} {\bibfnamefont {T.~C.}\ \bibnamefont {Ralph}},\ and\ \bibinfo
  {author} {\bibfnamefont {D.}~\bibnamefont {Branning}},\ }\bibfield  {title}
  {\bibinfo {title} {Demonstration of an all-optical quantum controlled-not
  gate},\ }\href {https://doi.org/10.1038/nature02054} {\bibfield  {journal}
  {\bibinfo  {journal} {Nature}\ }\textbf {\bibinfo {volume} {426}},\ \bibinfo
  {pages} {264} (\bibinfo {year} {2003})}\BibitemShut {NoStop}%
\bibitem [{\citenamefont {Franson}\ \emph {et~al.}(2002)\citenamefont
  {Franson}, \citenamefont {Donegan}, \citenamefont {Fitch}, \citenamefont
  {Jacobs},\ and\ \citenamefont {Pittman}}]{PhysRevLett.89.137901}%
  \BibitemOpen
  \bibfield  {author} {\bibinfo {author} {\bibfnamefont {J.~D.}\ \bibnamefont
  {Franson}}, \bibinfo {author} {\bibfnamefont {M.~M.}\ \bibnamefont
  {Donegan}}, \bibinfo {author} {\bibfnamefont {M.~J.}\ \bibnamefont {Fitch}},
  \bibinfo {author} {\bibfnamefont {B.~C.}\ \bibnamefont {Jacobs}},\ and\
  \bibinfo {author} {\bibfnamefont {T.~B.}\ \bibnamefont {Pittman}},\
  }\bibfield  {title} {\bibinfo {title} {High-fidelity quantum logic operations
  using linear optical elements},\ }\href
  {https://doi.org/10.1103/PhysRevLett.89.137901} {\bibfield  {journal}
  {\bibinfo  {journal} {Phys. Rev. Lett.}\ }\textbf {\bibinfo {volume} {89}},\
  \bibinfo {pages} {137901} (\bibinfo {year} {2002})}\BibitemShut {NoStop}%
\bibitem [{\citenamefont {Gasparoni}\ \emph {et~al.}(2004)\citenamefont
  {Gasparoni}, \citenamefont {Pan}, \citenamefont {Walther}, \citenamefont
  {Rudolph},\ and\ \citenamefont {Zeilinger}}]{PhysRevLett.93.020504}%
  \BibitemOpen
  \bibfield  {author} {\bibinfo {author} {\bibfnamefont {S.}~\bibnamefont
  {Gasparoni}}, \bibinfo {author} {\bibfnamefont {J.-W.}\ \bibnamefont {Pan}},
  \bibinfo {author} {\bibfnamefont {P.}~\bibnamefont {Walther}}, \bibinfo
  {author} {\bibfnamefont {T.}~\bibnamefont {Rudolph}},\ and\ \bibinfo {author}
  {\bibfnamefont {A.}~\bibnamefont {Zeilinger}},\ }\bibfield  {title} {\bibinfo
  {title} {Realization of a photonic controlled-not gate sufficient for quantum
  computation},\ }\href {https://doi.org/10.1103/PhysRevLett.93.020504}
  {\bibfield  {journal} {\bibinfo  {journal} {Phys. Rev. Lett.}\ }\textbf
  {\bibinfo {volume} {93}},\ \bibinfo {pages} {020504} (\bibinfo {year}
  {2004})}\BibitemShut {NoStop}%
\bibitem [{\citenamefont {Nasr}\ \emph {et~al.}(2003)\citenamefont {Nasr},
  \citenamefont {Saleh}, \citenamefont {Sergienko},\ and\ \citenamefont
  {Teich}}]{PhysRevLett.91.083601}%
  \BibitemOpen
  \bibfield  {author} {\bibinfo {author} {\bibfnamefont {M.~B.}\ \bibnamefont
  {Nasr}}, \bibinfo {author} {\bibfnamefont {B.~E.~A.}\ \bibnamefont {Saleh}},
  \bibinfo {author} {\bibfnamefont {A.~V.}\ \bibnamefont {Sergienko}},\ and\
  \bibinfo {author} {\bibfnamefont {M.~C.}\ \bibnamefont {Teich}},\ }\bibfield
  {title} {\bibinfo {title} {Demonstration of dispersion-canceled
  quantum-optical coherence tomography},\ }\href
  {https://doi.org/10.1103/PhysRevLett.91.083601} {\bibfield  {journal}
  {\bibinfo  {journal} {Phys. Rev. Lett.}\ }\textbf {\bibinfo {volume} {91}},\
  \bibinfo {pages} {083601} (\bibinfo {year} {2003})}\BibitemShut {NoStop}%
\bibitem [{\citenamefont {Lyons}\ \emph {et~al.}(2018)\citenamefont {Lyons},
  \citenamefont {Knee}, \citenamefont {Bolduc}, \citenamefont {Roger},
  \citenamefont {Leach}, \citenamefont {Gauger},\ and\ \citenamefont
  {Faccio}}]{Lyons2018}%
  \BibitemOpen
  \bibfield  {author} {\bibinfo {author} {\bibfnamefont {A.}~\bibnamefont
  {Lyons}}, \bibinfo {author} {\bibfnamefont {G.~C.}\ \bibnamefont {Knee}},
  \bibinfo {author} {\bibfnamefont {E.}~\bibnamefont {Bolduc}}, \bibinfo
  {author} {\bibfnamefont {T.}~\bibnamefont {Roger}}, \bibinfo {author}
  {\bibfnamefont {J.}~\bibnamefont {Leach}}, \bibinfo {author} {\bibfnamefont
  {E.~M.}\ \bibnamefont {Gauger}},\ and\ \bibinfo {author} {\bibfnamefont
  {D.}~\bibnamefont {Faccio}},\ }\bibfield  {title} {\bibinfo {title}
  {Attosecond-resolution {{Hong}}-{{Ou}}-{{Mandel}} interferometry},\ }\href
  {https://doi.org/10.1126/sciadv.aap9416} {\bibfield  {journal} {\bibinfo
  {journal} {Sci. Adv.}\ }\textbf {\bibinfo {volume} {4}},\ \bibinfo {pages}
  {eaap9416} (\bibinfo {year} {2018})}\BibitemShut {NoStop}%
\bibitem [{\citenamefont {Ricci}\ \emph
  {et~al.}(2004{\natexlab{a}})\citenamefont {Ricci}, \citenamefont {Martini},
  \citenamefont {Cerf}, \citenamefont {Filip}, \citenamefont
  {Fiur\'a\ifmmode~\check{s}\else \v{s}\fi{}ek},\ and\ \citenamefont
  {Macchiavello}}]{PhysRevLett.93.170501}%
  \BibitemOpen
  \bibfield  {author} {\bibinfo {author} {\bibfnamefont {M.}~\bibnamefont
  {Ricci}}, \bibinfo {author} {\bibfnamefont {F.~D.}\ \bibnamefont {Martini}},
  \bibinfo {author} {\bibfnamefont {N.~J.}\ \bibnamefont {Cerf}}, \bibinfo
  {author} {\bibfnamefont {R.}~\bibnamefont {Filip}}, \bibinfo {author}
  {\bibfnamefont {J.}~\bibnamefont {Fiur\'a\ifmmode~\check{s}\else
  \v{s}\fi{}ek}},\ and\ \bibinfo {author} {\bibfnamefont {C.}~\bibnamefont
  {Macchiavello}},\ }\bibfield  {title} {\bibinfo {title} {Experimental
  purification of single qubits},\ }\href
  {https://doi.org/10.1103/PhysRevLett.93.170501} {\bibfield  {journal}
  {\bibinfo  {journal} {Phys. Rev. Lett.}\ }\textbf {\bibinfo {volume} {93}},\
  \bibinfo {pages} {170501} (\bibinfo {year} {2004}{\natexlab{a}})}\BibitemShut
  {NoStop}%
\bibitem [{\citenamefont {Ricci}\ \emph
  {et~al.}(2004{\natexlab{b}})\citenamefont {Ricci}, \citenamefont {Sciarrino},
  \citenamefont {Sias},\ and\ \citenamefont
  {De~Martini}}]{PhysRevLett.92.047901}%
  \BibitemOpen
  \bibfield  {author} {\bibinfo {author} {\bibfnamefont {M.}~\bibnamefont
  {Ricci}}, \bibinfo {author} {\bibfnamefont {F.}~\bibnamefont {Sciarrino}},
  \bibinfo {author} {\bibfnamefont {C.}~\bibnamefont {Sias}},\ and\ \bibinfo
  {author} {\bibfnamefont {F.}~\bibnamefont {De~Martini}},\ }\bibfield  {title}
  {\bibinfo {title} {Teleportation scheme implementing the universal optimal
  quantum cloning machine and the universal not gate},\ }\href
  {https://doi.org/10.1103/PhysRevLett.92.047901} {\bibfield  {journal}
  {\bibinfo  {journal} {Phys. Rev. Lett.}\ }\textbf {\bibinfo {volume} {92}},\
  \bibinfo {pages} {047901} (\bibinfo {year} {2004}{\natexlab{b}})}\BibitemShut
  {NoStop}%
\bibitem [{\citenamefont {Nagata}\ \emph {et~al.}(2007)\citenamefont {Nagata},
  \citenamefont {Okamoto}, \citenamefont {O{\textquoteright}Brien},
  \citenamefont {Sasaki},\ and\ \citenamefont {Takeuchi}}]{Nagata726}%
  \BibitemOpen
  \bibfield  {author} {\bibinfo {author} {\bibfnamefont {T.}~\bibnamefont
  {Nagata}}, \bibinfo {author} {\bibfnamefont {R.}~\bibnamefont {Okamoto}},
  \bibinfo {author} {\bibfnamefont {J.~L.}\ \bibnamefont
  {O{\textquoteright}Brien}}, \bibinfo {author} {\bibfnamefont
  {K.}~\bibnamefont {Sasaki}},\ and\ \bibinfo {author} {\bibfnamefont
  {S.}~\bibnamefont {Takeuchi}},\ }\bibfield  {title} {\bibinfo {title}
  {Beating the standard quantum limit with four-entangled photons},\ }\href
  {https://doi.org/10.1126/science.1138007} {\bibfield  {journal} {\bibinfo
  {journal} {Science}\ }\textbf {\bibinfo {volume} {316}},\ \bibinfo {pages}
  {726} (\bibinfo {year} {2007})}\BibitemShut {NoStop}%
\bibitem [{\citenamefont {Agne}\ \emph {et~al.}(2017)\citenamefont {Agne},
  \citenamefont {Kauten}, \citenamefont {Jin}, \citenamefont {Meyer-Scott},
  \citenamefont {Salvail}, \citenamefont {Hamel}, \citenamefont {Resch},
  \citenamefont {Weihs},\ and\ \citenamefont
  {Jennewein}}]{PhysRevLett.118.153602}%
  \BibitemOpen
  \bibfield  {author} {\bibinfo {author} {\bibfnamefont {S.}~\bibnamefont
  {Agne}}, \bibinfo {author} {\bibfnamefont {T.}~\bibnamefont {Kauten}},
  \bibinfo {author} {\bibfnamefont {J.}~\bibnamefont {Jin}}, \bibinfo {author}
  {\bibfnamefont {E.}~\bibnamefont {Meyer-Scott}}, \bibinfo {author}
  {\bibfnamefont {J.~Z.}\ \bibnamefont {Salvail}}, \bibinfo {author}
  {\bibfnamefont {D.~R.}\ \bibnamefont {Hamel}}, \bibinfo {author}
  {\bibfnamefont {K.~J.}\ \bibnamefont {Resch}}, \bibinfo {author}
  {\bibfnamefont {G.}~\bibnamefont {Weihs}},\ and\ \bibinfo {author}
  {\bibfnamefont {T.}~\bibnamefont {Jennewein}},\ }\bibfield  {title} {\bibinfo
  {title} {Observation of genuine three-photon interference},\ }\href
  {https://doi.org/10.1103/PhysRevLett.118.153602} {\bibfield  {journal}
  {\bibinfo  {journal} {Phys. Rev. Lett.}\ }\textbf {\bibinfo {volume} {118}},\
  \bibinfo {pages} {153602} (\bibinfo {year} {2017})}\BibitemShut {NoStop}%
\bibitem [{\citenamefont {Menssen}\ \emph {et~al.}(2017)\citenamefont
  {Menssen}, \citenamefont {Jones}, \citenamefont {Metcalf}, \citenamefont
  {Tichy}, \citenamefont {Barz}, \citenamefont {Kolthammer},\ and\
  \citenamefont {Walmsley}}]{PhysRevLett.118.153603}%
  \BibitemOpen
  \bibfield  {author} {\bibinfo {author} {\bibfnamefont {A.~J.}\ \bibnamefont
  {Menssen}}, \bibinfo {author} {\bibfnamefont {A.~E.}\ \bibnamefont {Jones}},
  \bibinfo {author} {\bibfnamefont {B.~J.}\ \bibnamefont {Metcalf}}, \bibinfo
  {author} {\bibfnamefont {M.~C.}\ \bibnamefont {Tichy}}, \bibinfo {author}
  {\bibfnamefont {S.}~\bibnamefont {Barz}}, \bibinfo {author} {\bibfnamefont
  {W.~S.}\ \bibnamefont {Kolthammer}},\ and\ \bibinfo {author} {\bibfnamefont
  {I.~A.}\ \bibnamefont {Walmsley}},\ }\bibfield  {title} {\bibinfo {title}
  {Distinguishability and many-particle interference},\ }\href
  {https://doi.org/10.1103/PhysRevLett.118.153603} {\bibfield  {journal}
  {\bibinfo  {journal} {Phys. Rev. Lett.}\ }\textbf {\bibinfo {volume} {118}},\
  \bibinfo {pages} {153603} (\bibinfo {year} {2017})}\BibitemShut {NoStop}%
\bibitem [{\citenamefont {Aaronson}\ and\ \citenamefont
  {Arkhipov}(2010)}]{aaronson2010computational}%
  \BibitemOpen
  \bibfield  {author} {\bibinfo {author} {\bibfnamefont {S.}~\bibnamefont
  {Aaronson}}\ and\ \bibinfo {author} {\bibfnamefont {A.}~\bibnamefont
  {Arkhipov}},\ }\href@noop {} {\bibinfo {title} {The computational complexity
  of linear optics}} (\bibinfo {year} {2010}),\ \Eprint
  {https://arxiv.org/abs/1011.3245} {arXiv:1011.3245 [quant-ph]} \BibitemShut
  {NoStop}%
\bibitem [{\citenamefont {Broome}\ \emph {et~al.}(2013)\citenamefont {Broome},
  \citenamefont {Fedrizzi}, \citenamefont {Rahimi-Keshari}, \citenamefont
  {Dove}, \citenamefont {Aaronson}, \citenamefont {Ralph},\ and\ \citenamefont
  {White}}]{Broome794}%
  \BibitemOpen
  \bibfield  {author} {\bibinfo {author} {\bibfnamefont {M.~A.}\ \bibnamefont
  {Broome}}, \bibinfo {author} {\bibfnamefont {A.}~\bibnamefont {Fedrizzi}},
  \bibinfo {author} {\bibfnamefont {S.}~\bibnamefont {Rahimi-Keshari}},
  \bibinfo {author} {\bibfnamefont {J.}~\bibnamefont {Dove}}, \bibinfo {author}
  {\bibfnamefont {S.}~\bibnamefont {Aaronson}}, \bibinfo {author}
  {\bibfnamefont {T.~C.}\ \bibnamefont {Ralph}},\ and\ \bibinfo {author}
  {\bibfnamefont {A.~G.}\ \bibnamefont {White}},\ }\bibfield  {title} {\bibinfo
  {title} {Photonic boson sampling in a tunable circuit},\ }\href
  {https://doi.org/10.1126/science.1231440} {\bibfield  {journal} {\bibinfo
  {journal} {Science}\ }\textbf {\bibinfo {volume} {339}},\ \bibinfo {pages}
  {794} (\bibinfo {year} {2013})}\BibitemShut {NoStop}%
\bibitem [{\citenamefont {Bentivegna}\ \emph {et~al.}(2015)\citenamefont
  {Bentivegna}, \citenamefont {Spagnolo}, \citenamefont {Vitelli},
  \citenamefont {Flamini}, \citenamefont {Viggianiello}, \citenamefont
  {Latmiral}, \citenamefont {Mataloni}, \citenamefont {Brod}, \citenamefont
  {Galv{\~a}o}, \citenamefont {Crespi}, \citenamefont {Ramponi}, \citenamefont
  {Osellame},\ and\ \citenamefont {Sciarrino}}]{Bentivegnae1400255}%
  \BibitemOpen
  \bibfield  {author} {\bibinfo {author} {\bibfnamefont {M.}~\bibnamefont
  {Bentivegna}}, \bibinfo {author} {\bibfnamefont {N.}~\bibnamefont
  {Spagnolo}}, \bibinfo {author} {\bibfnamefont {C.}~\bibnamefont {Vitelli}},
  \bibinfo {author} {\bibfnamefont {F.}~\bibnamefont {Flamini}}, \bibinfo
  {author} {\bibfnamefont {N.}~\bibnamefont {Viggianiello}}, \bibinfo {author}
  {\bibfnamefont {L.}~\bibnamefont {Latmiral}}, \bibinfo {author}
  {\bibfnamefont {P.}~\bibnamefont {Mataloni}}, \bibinfo {author}
  {\bibfnamefont {D.~J.}\ \bibnamefont {Brod}}, \bibinfo {author}
  {\bibfnamefont {E.~F.}\ \bibnamefont {Galv{\~a}o}}, \bibinfo {author}
  {\bibfnamefont {A.}~\bibnamefont {Crespi}}, \bibinfo {author} {\bibfnamefont
  {R.}~\bibnamefont {Ramponi}}, \bibinfo {author} {\bibfnamefont
  {R.}~\bibnamefont {Osellame}},\ and\ \bibinfo {author} {\bibfnamefont
  {F.}~\bibnamefont {Sciarrino}},\ }\bibfield  {title} {\bibinfo {title}
  {Experimental scattershot boson sampling},\ }\href
  {https://doi.org/10.1126/sciadv.1400255} {\bibfield  {journal} {\bibinfo
  {journal} {Sci. Adv.}\ }\textbf {\bibinfo {volume} {1}},\ \bibinfo {pages}
  {e1400255} (\bibinfo {year} {2015})}\BibitemShut {NoStop}%
\bibitem [{\citenamefont {Stobi{\'n}ska}\ \emph {et~al.}(2019)\citenamefont
  {Stobi{\'n}ska}, \citenamefont {Buraczewski}, \citenamefont {Moore},
  \citenamefont {Clements}, \citenamefont {Renema}, \citenamefont {Nam},
  \citenamefont {Gerrits}, \citenamefont {Lita}, \citenamefont {Kolthammer},
  \citenamefont {Eckstein},\ and\ \citenamefont
  {Walmsley}}]{Stobinskaeaau9674}%
  \BibitemOpen
  \bibfield  {author} {\bibinfo {author} {\bibfnamefont {M.}~\bibnamefont
  {Stobi{\'n}ska}}, \bibinfo {author} {\bibfnamefont {A.}~\bibnamefont
  {Buraczewski}}, \bibinfo {author} {\bibfnamefont {M.}~\bibnamefont {Moore}},
  \bibinfo {author} {\bibfnamefont {W.~R.}\ \bibnamefont {Clements}}, \bibinfo
  {author} {\bibfnamefont {J.~J.}\ \bibnamefont {Renema}}, \bibinfo {author}
  {\bibfnamefont {S.~W.}\ \bibnamefont {Nam}}, \bibinfo {author} {\bibfnamefont
  {T.}~\bibnamefont {Gerrits}}, \bibinfo {author} {\bibfnamefont
  {A.}~\bibnamefont {Lita}}, \bibinfo {author} {\bibfnamefont {W.~S.}\
  \bibnamefont {Kolthammer}}, \bibinfo {author} {\bibfnamefont
  {A.}~\bibnamefont {Eckstein}},\ and\ \bibinfo {author} {\bibfnamefont
  {I.~A.}\ \bibnamefont {Walmsley}},\ }\bibfield  {title} {\bibinfo {title}
  {Quantum interference enables constant-time quantum information processing},\
  }\href {https://doi.org/10.1126/sciadv.aau9674} {\bibfield  {journal}
  {\bibinfo  {journal} {Sci. Adv.}\ }\textbf {\bibinfo {volume} {5}},\ \bibinfo
  {pages} {eaau9674} (\bibinfo {year} {2019})}\BibitemShut {NoStop}%
\bibitem [{\citenamefont {Zhang}\ \emph
  {et~al.}(2016{\natexlab{a}})\citenamefont {Zhang}, \citenamefont {Roux},
  \citenamefont {Konrad}, \citenamefont {Agnew}, \citenamefont {Leach},\ and\
  \citenamefont {Forbes}}]{Zhange1501165}%
  \BibitemOpen
  \bibfield  {author} {\bibinfo {author} {\bibfnamefont {Y.}~\bibnamefont
  {Zhang}}, \bibinfo {author} {\bibfnamefont {F.~S.}\ \bibnamefont {Roux}},
  \bibinfo {author} {\bibfnamefont {T.}~\bibnamefont {Konrad}}, \bibinfo
  {author} {\bibfnamefont {M.}~\bibnamefont {Agnew}}, \bibinfo {author}
  {\bibfnamefont {J.}~\bibnamefont {Leach}},\ and\ \bibinfo {author}
  {\bibfnamefont {A.}~\bibnamefont {Forbes}},\ }\bibfield  {title} {\bibinfo
  {title} {Engineering two-photon high-dimensional states through quantum
  interference},\ }\href {https://doi.org/10.1126/sciadv.1501165} {\bibfield
  {journal} {\bibinfo  {journal} {Sci. Adv.}\ }\textbf {\bibinfo {volume}
  {2}},\ \bibinfo {pages} {e1501165} (\bibinfo {year}
  {2016}{\natexlab{a}})}\BibitemShut {NoStop}%
\bibitem [{\citenamefont {Tischler}\ \emph {et~al.}(2015)\citenamefont
  {Tischler}, \citenamefont {B\"use}, \citenamefont {Helt}, \citenamefont
  {Juan}, \citenamefont {Piro}, \citenamefont {Ghosh}, \citenamefont {Steel},\
  and\ \citenamefont {Molina-Terriza}}]{PhysRevLett.115.193602}%
  \BibitemOpen
  \bibfield  {author} {\bibinfo {author} {\bibfnamefont {N.}~\bibnamefont
  {Tischler}}, \bibinfo {author} {\bibfnamefont {A.}~\bibnamefont {B\"use}},
  \bibinfo {author} {\bibfnamefont {L.~G.}\ \bibnamefont {Helt}}, \bibinfo
  {author} {\bibfnamefont {M.~L.}\ \bibnamefont {Juan}}, \bibinfo {author}
  {\bibfnamefont {N.}~\bibnamefont {Piro}}, \bibinfo {author} {\bibfnamefont
  {J.}~\bibnamefont {Ghosh}}, \bibinfo {author} {\bibfnamefont {M.~J.}\
  \bibnamefont {Steel}},\ and\ \bibinfo {author} {\bibfnamefont
  {G.}~\bibnamefont {Molina-Terriza}},\ }\bibfield  {title} {\bibinfo {title}
  {Measurement and shaping of biphoton spectral wave functions},\ }\href
  {https://doi.org/10.1103/PhysRevLett.115.193602} {\bibfield  {journal}
  {\bibinfo  {journal} {Phys. Rev. Lett.}\ }\textbf {\bibinfo {volume} {115}},\
  \bibinfo {pages} {193602} (\bibinfo {year} {2015})}\BibitemShut {NoStop}%
\bibitem [{\citenamefont {Law}\ \emph {et~al.}(2000)\citenamefont {Law},
  \citenamefont {Walmsley},\ and\ \citenamefont
  {Eberly}}]{PhysRevLett.84.5304}%
  \BibitemOpen
  \bibfield  {author} {\bibinfo {author} {\bibfnamefont {C.~K.}\ \bibnamefont
  {Law}}, \bibinfo {author} {\bibfnamefont {I.~A.}\ \bibnamefont {Walmsley}},\
  and\ \bibinfo {author} {\bibfnamefont {J.~H.}\ \bibnamefont {Eberly}},\
  }\bibfield  {title} {\bibinfo {title} {Continuous frequency entanglement:
  Effective finite hilbert space and entropy control},\ }\href
  {https://doi.org/10.1103/PhysRevLett.84.5304} {\bibfield  {journal} {\bibinfo
   {journal} {Phys. Rev. Lett.}\ }\textbf {\bibinfo {volume} {84}},\ \bibinfo
  {pages} {5304} (\bibinfo {year} {2000})}\BibitemShut {NoStop}%
\bibitem [{\citenamefont {Mosley}\ \emph {et~al.}(2008)\citenamefont {Mosley},
  \citenamefont {Lundeen}, \citenamefont {Smith}, \citenamefont {Wasylczyk},
  \citenamefont {U'Ren}, \citenamefont {Silberhorn},\ and\ \citenamefont
  {Walmsley}}]{PhysRevLett.100.133601}%
  \BibitemOpen
  \bibfield  {author} {\bibinfo {author} {\bibfnamefont {P.~J.}\ \bibnamefont
  {Mosley}}, \bibinfo {author} {\bibfnamefont {J.~S.}\ \bibnamefont {Lundeen}},
  \bibinfo {author} {\bibfnamefont {B.~J.}\ \bibnamefont {Smith}}, \bibinfo
  {author} {\bibfnamefont {P.}~\bibnamefont {Wasylczyk}}, \bibinfo {author}
  {\bibfnamefont {A.~B.}\ \bibnamefont {U'Ren}}, \bibinfo {author}
  {\bibfnamefont {C.}~\bibnamefont {Silberhorn}},\ and\ \bibinfo {author}
  {\bibfnamefont {I.~A.}\ \bibnamefont {Walmsley}},\ }\bibfield  {title}
  {\bibinfo {title} {Heralded generation of ultrafast single photons in pure
  quantum states},\ }\href {https://doi.org/10.1103/PhysRevLett.100.133601}
  {\bibfield  {journal} {\bibinfo  {journal} {Phys. Rev. Lett.}\ }\textbf
  {\bibinfo {volume} {100}},\ \bibinfo {pages} {133601} (\bibinfo {year}
  {2008})}\BibitemShut {NoStop}%
\bibitem [{\citenamefont {Santori}\ \emph {et~al.}(2002)\citenamefont
  {Santori}, \citenamefont {Fattal}, \citenamefont {Vu{\v c}kovi{\'c}},
  \citenamefont {Solomon},\ and\ \citenamefont {Yamamoto}}]{Santori2002}%
  \BibitemOpen
  \bibfield  {author} {\bibinfo {author} {\bibfnamefont {C.}~\bibnamefont
  {Santori}}, \bibinfo {author} {\bibfnamefont {D.}~\bibnamefont {Fattal}},
  \bibinfo {author} {\bibfnamefont {J.}~\bibnamefont {Vu{\v c}kovi{\'c}}},
  \bibinfo {author} {\bibfnamefont {G.~S.}\ \bibnamefont {Solomon}},\ and\
  \bibinfo {author} {\bibfnamefont {Y.}~\bibnamefont {Yamamoto}},\ }\bibfield
  {title} {\bibinfo {title} {Indistinguishable photons from a single-photon
  device},\ }\href {https://doi.org/10.1038/nature01086} {\bibfield  {journal}
  {\bibinfo  {journal} {Nature}\ }\textbf {\bibinfo {volume} {419}},\ \bibinfo
  {pages} {594} (\bibinfo {year} {2002})}\BibitemShut {NoStop}%
\bibitem [{\citenamefont {Cohen}\ \emph {et~al.}(2009)\citenamefont {Cohen},
  \citenamefont {Lundeen}, \citenamefont {Smith}, \citenamefont {Puentes},
  \citenamefont {Mosley},\ and\ \citenamefont
  {Walmsley}}]{PhysRevLett.102.123603}%
  \BibitemOpen
  \bibfield  {author} {\bibinfo {author} {\bibfnamefont {O.}~\bibnamefont
  {Cohen}}, \bibinfo {author} {\bibfnamefont {J.~S.}\ \bibnamefont {Lundeen}},
  \bibinfo {author} {\bibfnamefont {B.~J.}\ \bibnamefont {Smith}}, \bibinfo
  {author} {\bibfnamefont {G.}~\bibnamefont {Puentes}}, \bibinfo {author}
  {\bibfnamefont {P.~J.}\ \bibnamefont {Mosley}},\ and\ \bibinfo {author}
  {\bibfnamefont {I.~A.}\ \bibnamefont {Walmsley}},\ }\bibfield  {title}
  {\bibinfo {title} {Tailored photon-pair generation in optical fibers},\
  }\href {https://doi.org/10.1103/PhysRevLett.102.123603} {\bibfield  {journal}
  {\bibinfo  {journal} {Phys. Rev. Lett.}\ }\textbf {\bibinfo {volume} {102}},\
  \bibinfo {pages} {123603} (\bibinfo {year} {2009})}\BibitemShut {NoStop}%
\bibitem [{\citenamefont {Vergyris}\ \emph {et~al.}(2016)\citenamefont
  {Vergyris}, \citenamefont {Meany}, \citenamefont {Lunghi}, \citenamefont
  {Sauder}, \citenamefont {Downes}, \citenamefont {Steel}, \citenamefont
  {Withford}, \citenamefont {Alibart},\ and\ \citenamefont
  {Tanzilli}}]{tanzilli}%
  \BibitemOpen
  \bibfield  {author} {\bibinfo {author} {\bibfnamefont {P.}~\bibnamefont
  {Vergyris}}, \bibinfo {author} {\bibfnamefont {T.}~\bibnamefont {Meany}},
  \bibinfo {author} {\bibfnamefont {T.}~\bibnamefont {Lunghi}}, \bibinfo
  {author} {\bibfnamefont {G.}~\bibnamefont {Sauder}}, \bibinfo {author}
  {\bibfnamefont {J.}~\bibnamefont {Downes}}, \bibinfo {author} {\bibfnamefont
  {M.~J.}\ \bibnamefont {Steel}}, \bibinfo {author} {\bibfnamefont {M.~J.}\
  \bibnamefont {Withford}}, \bibinfo {author} {\bibfnamefont {O.}~\bibnamefont
  {Alibart}},\ and\ \bibinfo {author} {\bibfnamefont {S.}~\bibnamefont
  {Tanzilli}},\ }\bibfield  {title} {\bibinfo {title} {On-chip generation of
  heralded photon-number states},\ }\href {https://doi.org/10.1038/srep35975}
  {\bibfield  {journal} {\bibinfo  {journal} {Sci. Rep.}\ }\textbf {\bibinfo
  {volume} {6}},\ \bibinfo {pages} {35975} (\bibinfo {year}
  {2016})}\BibitemShut {NoStop}%
\bibitem [{\citenamefont {Ollivier}\ \emph {et~al.}(2021)\citenamefont
  {Ollivier}, \citenamefont {Thomas}, \citenamefont {Wein}, \citenamefont
  {de~Buy~Wenniger}, \citenamefont {Coste}, \citenamefont {Loredo},
  \citenamefont {Somaschi}, \citenamefont {Harouri}, \citenamefont {Lemaitre},
  \citenamefont {Sagnes}, \citenamefont {Lanco}, \citenamefont {Simon},
  \citenamefont {Anton}, \citenamefont {Krebs},\ and\ \citenamefont
  {Senellart}}]{PhysRevLett.126.063602}%
  \BibitemOpen
  \bibfield  {author} {\bibinfo {author} {\bibfnamefont {H.}~\bibnamefont
  {Ollivier}}, \bibinfo {author} {\bibfnamefont {S.~E.}\ \bibnamefont
  {Thomas}}, \bibinfo {author} {\bibfnamefont {S.~C.}\ \bibnamefont {Wein}},
  \bibinfo {author} {\bibfnamefont {I.~M.}\ \bibnamefont {de~Buy~Wenniger}},
  \bibinfo {author} {\bibfnamefont {N.}~\bibnamefont {Coste}}, \bibinfo
  {author} {\bibfnamefont {J.~C.}\ \bibnamefont {Loredo}}, \bibinfo {author}
  {\bibfnamefont {N.}~\bibnamefont {Somaschi}}, \bibinfo {author}
  {\bibfnamefont {A.}~\bibnamefont {Harouri}}, \bibinfo {author} {\bibfnamefont
  {A.}~\bibnamefont {Lemaitre}}, \bibinfo {author} {\bibfnamefont
  {I.}~\bibnamefont {Sagnes}}, \bibinfo {author} {\bibfnamefont
  {L.}~\bibnamefont {Lanco}}, \bibinfo {author} {\bibfnamefont
  {C.}~\bibnamefont {Simon}}, \bibinfo {author} {\bibfnamefont
  {C.}~\bibnamefont {Anton}}, \bibinfo {author} {\bibfnamefont
  {O.}~\bibnamefont {Krebs}},\ and\ \bibinfo {author} {\bibfnamefont
  {P.}~\bibnamefont {Senellart}},\ }\bibfield  {title} {\bibinfo {title}
  {Hong-ou-mandel interference with imperfect single photon sources},\ }\href
  {https://doi.org/10.1103/PhysRevLett.126.063602} {\bibfield  {journal}
  {\bibinfo  {journal} {Phys. Rev. Lett.}\ }\textbf {\bibinfo {volume} {126}},\
  \bibinfo {pages} {063602} (\bibinfo {year} {2021})}\BibitemShut {NoStop}%
\bibitem [{\citenamefont {Senellart}\ \emph {et~al.}(2017)\citenamefont
  {Senellart}, \citenamefont {Solomon},\ and\ \citenamefont
  {White}}]{Senellart2017}%
  \BibitemOpen
  \bibfield  {author} {\bibinfo {author} {\bibfnamefont {P.}~\bibnamefont
  {Senellart}}, \bibinfo {author} {\bibfnamefont {G.}~\bibnamefont {Solomon}},\
  and\ \bibinfo {author} {\bibfnamefont {A.}~\bibnamefont {White}},\ }\bibfield
   {title} {\bibinfo {title} {High-performance semiconductor quantum-dot
  single-photon sources},\ }\href {https://doi.org/10.1038/nnano.2017.218}
  {\bibfield  {journal} {\bibinfo  {journal} {Nat. Nanotechnol.}\ }\textbf
  {\bibinfo {volume} {12}},\ \bibinfo {pages} {1026} (\bibinfo {year}
  {2017})}\BibitemShut {NoStop}%
\bibitem [{\citenamefont {Francesconi}\ \emph {et~al.}(2020)\citenamefont
  {Francesconi}, \citenamefont {Baboux}, \citenamefont {Raymond}, \citenamefont
  {Fabre}, \citenamefont {Boucher}, \citenamefont {Lema\^{i}tre}, \citenamefont
  {Milman}, \citenamefont {Amanti},\ and\ \citenamefont
  {Ducci}}]{Francesconi:20}%
  \BibitemOpen
  \bibfield  {author} {\bibinfo {author} {\bibfnamefont {S.}~\bibnamefont
  {Francesconi}}, \bibinfo {author} {\bibfnamefont {F.}~\bibnamefont {Baboux}},
  \bibinfo {author} {\bibfnamefont {A.}~\bibnamefont {Raymond}}, \bibinfo
  {author} {\bibfnamefont {N.}~\bibnamefont {Fabre}}, \bibinfo {author}
  {\bibfnamefont {G.}~\bibnamefont {Boucher}}, \bibinfo {author} {\bibfnamefont
  {A.}~\bibnamefont {Lema\^{i}tre}}, \bibinfo {author} {\bibfnamefont
  {P.}~\bibnamefont {Milman}}, \bibinfo {author} {\bibfnamefont {M.~I.}\
  \bibnamefont {Amanti}},\ and\ \bibinfo {author} {\bibfnamefont
  {S.}~\bibnamefont {Ducci}},\ }\bibfield  {title} {\bibinfo {title}
  {Engineering two-photon wavefunction and exchange statistics in a
  semiconductor chip},\ }\href {https://doi.org/10.1364/OPTICA.379477}
  {\bibfield  {journal} {\bibinfo  {journal} {Optica}\ }\textbf {\bibinfo
  {volume} {7}},\ \bibinfo {pages} {316} (\bibinfo {year} {2020})}\BibitemShut
  {NoStop}%
\bibitem [{\citenamefont {Zhang}\ \emph
  {et~al.}(2016{\natexlab{b}})\citenamefont {Zhang}, \citenamefont {Jiang},
  \citenamefont {Bell}, \citenamefont {Choi}, \citenamefont {Chae},\ and\
  \citenamefont {Xiong}}]{technologies4030025}%
  \BibitemOpen
  \bibfield  {author} {\bibinfo {author} {\bibfnamefont {X.}~\bibnamefont
  {Zhang}}, \bibinfo {author} {\bibfnamefont {R.}~\bibnamefont {Jiang}},
  \bibinfo {author} {\bibfnamefont {B.~A.}\ \bibnamefont {Bell}}, \bibinfo
  {author} {\bibfnamefont {D.-Y.}\ \bibnamefont {Choi}}, \bibinfo {author}
  {\bibfnamefont {C.~J.}\ \bibnamefont {Chae}},\ and\ \bibinfo {author}
  {\bibfnamefont {C.}~\bibnamefont {Xiong}},\ }\bibfield  {title} {\bibinfo
  {title} {Interfering {{Heralded Single Photons}} from {{Two Separate Silicon
  Nanowires Pumped}} at {{Different Wavelengths}}},\ }\href
  {https://doi.org/10.3390/technologies4030025} {\bibfield  {journal} {\bibinfo
   {journal} {Technologies}\ }\textbf {\bibinfo {volume} {4}},\ \bibinfo
  {pages} {25} (\bibinfo {year} {2016}{\natexlab{b}})}\BibitemShut {NoStop}%
\bibitem [{\citenamefont {Dorfman}\ and\ \citenamefont
  {Mukamel}(2014)}]{Dorfman2014}%
  \BibitemOpen
  \bibfield  {author} {\bibinfo {author} {\bibfnamefont {K.~E.}\ \bibnamefont
  {Dorfman}}\ and\ \bibinfo {author} {\bibfnamefont {S.}~\bibnamefont
  {Mukamel}},\ }\bibfield  {title} {\bibinfo {title} {Indistinguishability and
  correlations of photons generated by quantum emitters undergoing spectral
  diffusion},\ }\href {https://doi.org/10.1038/srep03996} {\bibfield  {journal}
  {\bibinfo  {journal} {Sci. Rep.}\ }\textbf {\bibinfo {volume} {4}},\ \bibinfo
  {pages} {3996} (\bibinfo {year} {2014})}\BibitemShut {NoStop}%
\bibitem [{\citenamefont {Hua}\ \emph {et~al.}(2021)\citenamefont {Hua},
  \citenamefont {Lunghi}, \citenamefont {Doutre}, \citenamefont {Vergyris},
  \citenamefont {Sauder}, \citenamefont {Charlier}, \citenamefont
  {Labont\'{e}}, \citenamefont {D'Auria}, \citenamefont {Martin}, \citenamefont
  {Tascu}, \citenamefont {Micheli}, \citenamefont {Tanzilli},\ and\
  \citenamefont {Alibart}}]{Hua:21}%
  \BibitemOpen
  \bibfield  {author} {\bibinfo {author} {\bibfnamefont {X.}~\bibnamefont
  {Hua}}, \bibinfo {author} {\bibfnamefont {T.}~\bibnamefont {Lunghi}},
  \bibinfo {author} {\bibfnamefont {F.}~\bibnamefont {Doutre}}, \bibinfo
  {author} {\bibfnamefont {P.}~\bibnamefont {Vergyris}}, \bibinfo {author}
  {\bibfnamefont {G.}~\bibnamefont {Sauder}}, \bibinfo {author} {\bibfnamefont
  {P.}~\bibnamefont {Charlier}}, \bibinfo {author} {\bibfnamefont
  {L.}~\bibnamefont {Labont\'{e}}}, \bibinfo {author} {\bibfnamefont
  {V.}~\bibnamefont {D'Auria}}, \bibinfo {author} {\bibfnamefont
  {A.}~\bibnamefont {Martin}}, \bibinfo {author} {\bibfnamefont
  {S.}~\bibnamefont {Tascu}}, \bibinfo {author} {\bibfnamefont {M.~P.~D.}\
  \bibnamefont {Micheli}}, \bibinfo {author} {\bibfnamefont {S.}~\bibnamefont
  {Tanzilli}},\ and\ \bibinfo {author} {\bibfnamefont {O.}~\bibnamefont
  {Alibart}},\ }\bibfield  {title} {\bibinfo {title} {Configurable heralded
  two-photon fock-states on a chip},\ }\href
  {https://doi.org/10.1364/OE.403552} {\bibfield  {journal} {\bibinfo
  {journal} {Opt. Express}\ }\textbf {\bibinfo {volume} {29}},\ \bibinfo
  {pages} {415} (\bibinfo {year} {2021})}\BibitemShut {NoStop}%
\bibitem [{\citenamefont {Legero}\ \emph {et~al.}(2004)\citenamefont {Legero},
  \citenamefont {Wilk}, \citenamefont {Hennrich}, \citenamefont {Rempe},\ and\
  \citenamefont {Kuhn}}]{PhysRevLett.93.070503}%
  \BibitemOpen
  \bibfield  {author} {\bibinfo {author} {\bibfnamefont {T.}~\bibnamefont
  {Legero}}, \bibinfo {author} {\bibfnamefont {T.}~\bibnamefont {Wilk}},
  \bibinfo {author} {\bibfnamefont {M.}~\bibnamefont {Hennrich}}, \bibinfo
  {author} {\bibfnamefont {G.}~\bibnamefont {Rempe}},\ and\ \bibinfo {author}
  {\bibfnamefont {A.}~\bibnamefont {Kuhn}},\ }\bibfield  {title} {\bibinfo
  {title} {Quantum beat of two single photons},\ }\href
  {https://doi.org/10.1103/PhysRevLett.93.070503} {\bibfield  {journal}
  {\bibinfo  {journal} {Phys. Rev. Lett.}\ }\textbf {\bibinfo {volume} {93}},\
  \bibinfo {pages} {070503} (\bibinfo {year} {2004})}\BibitemShut {NoStop}%
\bibitem [{\citenamefont {Patel}\ \emph {et~al.}(2010)\citenamefont {Patel},
  \citenamefont {Bennett}, \citenamefont {Farrer}, \citenamefont {Nicoll},
  \citenamefont {Ritchie},\ and\ \citenamefont {Shields}}]{Patel2010}%
  \BibitemOpen
  \bibfield  {author} {\bibinfo {author} {\bibfnamefont {R.~B.}\ \bibnamefont
  {Patel}}, \bibinfo {author} {\bibfnamefont {A.~J.}\ \bibnamefont {Bennett}},
  \bibinfo {author} {\bibfnamefont {I.}~\bibnamefont {Farrer}}, \bibinfo
  {author} {\bibfnamefont {C.~A.}\ \bibnamefont {Nicoll}}, \bibinfo {author}
  {\bibfnamefont {D.~A.}\ \bibnamefont {Ritchie}},\ and\ \bibinfo {author}
  {\bibfnamefont {A.~J.}\ \bibnamefont {Shields}},\ }\bibfield  {title}
  {\bibinfo {title} {Two-photon interference of the emission from electrically
  tunable remote quantum dots},\ }\href
  {https://doi.org/10.1038/nphoton.2010.161} {\bibfield  {journal} {\bibinfo
  {journal} {Nat. Photonics}\ }\textbf {\bibinfo {volume} {4}},\ \bibinfo
  {pages} {632} (\bibinfo {year} {2010})}\BibitemShut {NoStop}%
\bibitem [{\citenamefont {Lipka}\ and\ \citenamefont
  {Parniak}(2021)}]{Lipka2021a}%
  \BibitemOpen
  \bibfield  {author} {\bibinfo {author} {\bibfnamefont {M.}~\bibnamefont
  {Lipka}}\ and\ \bibinfo {author} {\bibfnamefont {M.}~\bibnamefont
  {Parniak}},\ }\href@noop {} {\bibinfo {title} {Single-photon hologram of a
  zero-area pulse}} (\bibinfo {year} {2021}),\ \Eprint
  {https://arxiv.org/abs/2105.02795} {arXiv:2105.02795 [quant-ph]} \BibitemShut
  {NoStop}%
\bibitem [{\citenamefont {Fedrizzi}\ \emph {et~al.}(2009)\citenamefont
  {Fedrizzi}, \citenamefont {Herbst}, \citenamefont {Aspelmeyer}, \citenamefont
  {Barbieri}, \citenamefont {Jennewein},\ and\ \citenamefont
  {Zeilinger}}]{Fedrizzi_2009}%
  \BibitemOpen
  \bibfield  {author} {\bibinfo {author} {\bibfnamefont {A.}~\bibnamefont
  {Fedrizzi}}, \bibinfo {author} {\bibfnamefont {T.}~\bibnamefont {Herbst}},
  \bibinfo {author} {\bibfnamefont {M.}~\bibnamefont {Aspelmeyer}}, \bibinfo
  {author} {\bibfnamefont {M.}~\bibnamefont {Barbieri}}, \bibinfo {author}
  {\bibfnamefont {T.}~\bibnamefont {Jennewein}},\ and\ \bibinfo {author}
  {\bibfnamefont {A.}~\bibnamefont {Zeilinger}},\ }\bibfield  {title} {\bibinfo
  {title} {Anti-symmetrization reveals hidden entanglement},\ }\href
  {https://doi.org/10.1088/1367-2630/11/10/103052} {\bibfield  {journal}
  {\bibinfo  {journal} {New J. Phys.}\ }\textbf {\bibinfo {volume} {11}},\
  \bibinfo {pages} {103052} (\bibinfo {year} {2009})}\BibitemShut {NoStop}%
\bibitem [{\citenamefont {Barbieri}\ \emph {et~al.}(2017)\citenamefont
  {Barbieri}, \citenamefont {Roccia}, \citenamefont {Mancino}, \citenamefont
  {Sbroscia}, \citenamefont {Gianani},\ and\ \citenamefont
  {Sciarrino}}]{scirepHOM}%
  \BibitemOpen
  \bibfield  {author} {\bibinfo {author} {\bibfnamefont {M.}~\bibnamefont
  {Barbieri}}, \bibinfo {author} {\bibfnamefont {E.}~\bibnamefont {Roccia}},
  \bibinfo {author} {\bibfnamefont {L.}~\bibnamefont {Mancino}}, \bibinfo
  {author} {\bibfnamefont {M.}~\bibnamefont {Sbroscia}}, \bibinfo {author}
  {\bibfnamefont {I.}~\bibnamefont {Gianani}},\ and\ \bibinfo {author}
  {\bibfnamefont {F.}~\bibnamefont {Sciarrino}},\ }\bibfield  {title} {\bibinfo
  {title} {{What Hong-Ou-Mandel interference says on two-photon frequency
  entanglement}},\ }\href {https://doi.org/10.1038/s41598-017-07555-4}
  {\bibfield  {journal} {\bibinfo  {journal} {Sci. Rep.}\ }\textbf {\bibinfo
  {volume} {7}},\ \bibinfo {pages} {7247} (\bibinfo {year} {2017})}\BibitemShut
  {NoStop}%
\bibitem [{\citenamefont {Faruque}\ \emph {et~al.}(2019)\citenamefont
  {Faruque}, \citenamefont {Sinclair}, \citenamefont {Bonneau}, \citenamefont
  {Ono}, \citenamefont {Silberhorn}, \citenamefont {Thompson},\ and\
  \citenamefont {Rarity}}]{PhysRevApplied.12.054029}%
  \BibitemOpen
  \bibfield  {author} {\bibinfo {author} {\bibfnamefont {I.~I.}\ \bibnamefont
  {Faruque}}, \bibinfo {author} {\bibfnamefont {G.~F.}\ \bibnamefont
  {Sinclair}}, \bibinfo {author} {\bibfnamefont {D.}~\bibnamefont {Bonneau}},
  \bibinfo {author} {\bibfnamefont {T.}~\bibnamefont {Ono}}, \bibinfo {author}
  {\bibfnamefont {C.}~\bibnamefont {Silberhorn}}, \bibinfo {author}
  {\bibfnamefont {M.~G.}\ \bibnamefont {Thompson}},\ and\ \bibinfo {author}
  {\bibfnamefont {J.~G.}\ \bibnamefont {Rarity}},\ }\bibfield  {title}
  {\bibinfo {title} {Estimating the indistinguishability of heralded single
  photons using second-order correlation},\ }\href
  {https://doi.org/10.1103/PhysRevApplied.12.054029} {\bibfield  {journal}
  {\bibinfo  {journal} {Phys. Rev. Applied}\ }\textbf {\bibinfo {volume}
  {12}},\ \bibinfo {pages} {054029} (\bibinfo {year} {2019})}\BibitemShut
  {NoStop}%
\bibitem [{\citenamefont {Bickel}\ \emph {et~al.}(1993)\citenamefont {Bickel},
  \citenamefont {Klaassen}, \citenamefont {Ritov},\ and\ \citenamefont
  {Wellner}}]{bickel1998efficient}%
  \BibitemOpen
  \bibfield  {author} {\bibinfo {author} {\bibfnamefont {P.~J.}\ \bibnamefont
  {Bickel}}, \bibinfo {author} {\bibfnamefont {C.~A.~J.}\ \bibnamefont
  {Klaassen}}, \bibinfo {author} {\bibfnamefont {Y.}~\bibnamefont {Ritov}},\
  and\ \bibinfo {author} {\bibfnamefont {J.~A.}\ \bibnamefont {Wellner}},\
  }\href@noop {} {\emph {\bibinfo {title} {Efficient and {{Adaptive
  Estimation}} for {{Semiparametric Models}}}}}\ (\bibinfo  {publisher}
  {{Springer}},\ \bibinfo {address} {{New York}},\ \bibinfo {year}
  {1993})\BibitemShut {NoStop}%
\bibitem [{\citenamefont {Tsiatis}(2006)}]{Tsiatis2006}%
  \BibitemOpen
  \bibfield  {author} {\bibinfo {author} {\bibfnamefont {A.}~\bibnamefont
  {Tsiatis}},\ }\href {https://doi.org/10.1007/0-387-37345-4} {\emph {\bibinfo
  {title} {Semiparametric {{Theory}} and {{Missing Data}}}}}\ (\bibinfo
  {publisher} {{Springer}},\ \bibinfo {address} {{New York}},\ \bibinfo {year}
  {2006})\BibitemShut {NoStop}%
\bibitem [{\citenamefont {Tsang}(2019)}]{PhysRevResearch.1.033006}%
  \BibitemOpen
  \bibfield  {author} {\bibinfo {author} {\bibfnamefont {M.}~\bibnamefont
  {Tsang}},\ }\bibfield  {title} {\bibinfo {title} {Semiparametric estimation
  for incoherent optical imaging},\ }\href
  {https://doi.org/10.1103/PhysRevResearch.1.033006} {\bibfield  {journal}
  {\bibinfo  {journal} {Phys. Rev. Research}\ }\textbf {\bibinfo {volume}
  {1}},\ \bibinfo {pages} {033006} (\bibinfo {year} {2019})}\BibitemShut
  {NoStop}%
\bibitem [{\citenamefont {Tsang}(2021)}]{Tsang2020b}%
  \BibitemOpen
  \bibfield  {author} {\bibinfo {author} {\bibfnamefont {M.}~\bibnamefont
  {Tsang}},\ }\href@noop {} {\bibinfo {title} {Quantum limit to subdiffraction
  incoherent optical imaging. {{II}}. {{A}} parametric-submodel approach}}
  (\bibinfo {year} {2021}),\ \Eprint {https://arxiv.org/abs/2010.03518v3}
  {arXiv:2010.03518v3 [quant-ph]} \BibitemShut {NoStop}%
\bibitem [{\citenamefont {Tsang}\ \emph {et~al.}(2020)\citenamefont {Tsang},
  \citenamefont {Albarelli},\ and\ \citenamefont {Datta}}]{PhysRevX.10.031023}%
  \BibitemOpen
  \bibfield  {author} {\bibinfo {author} {\bibfnamefont {M.}~\bibnamefont
  {Tsang}}, \bibinfo {author} {\bibfnamefont {F.}~\bibnamefont {Albarelli}},\
  and\ \bibinfo {author} {\bibfnamefont {A.}~\bibnamefont {Datta}},\ }\bibfield
   {title} {\bibinfo {title} {Quantum semiparametric estimation},\ }\href
  {https://doi.org/10.1103/PhysRevX.10.031023} {\bibfield  {journal} {\bibinfo
  {journal} {Phys. Rev. X}\ }\textbf {\bibinfo {volume} {10}},\ \bibinfo
  {pages} {031023} (\bibinfo {year} {2020})}\BibitemShut {NoStop}%
\bibitem [{\citenamefont {Suzuki}\ \emph {et~al.}(2020)\citenamefont {Suzuki},
  \citenamefont {Yang},\ and\ \citenamefont {Hayashi}}]{Suzuki2019a}%
  \BibitemOpen
  \bibfield  {author} {\bibinfo {author} {\bibfnamefont {J.}~\bibnamefont
  {Suzuki}}, \bibinfo {author} {\bibfnamefont {Y.}~\bibnamefont {Yang}},\ and\
  \bibinfo {author} {\bibfnamefont {M.}~\bibnamefont {Hayashi}},\ }\bibfield
  {title} {\bibinfo {title} {Quantum state estimation with nuisance
  parameters},\ }\href {https://doi.org/10.1088/1751-8121/ab8b78} {\bibfield
  {journal} {\bibinfo  {journal} {J. Phys. A}\ }\textbf {\bibinfo {volume}
  {53}},\ \bibinfo {pages} {453001} (\bibinfo {year} {2020})}\BibitemShut
  {NoStop}%
\bibitem [{\citenamefont {Suzuki}(2020)}]{Suzuki2019}%
  \BibitemOpen
  \bibfield  {author} {\bibinfo {author} {\bibfnamefont {J.}~\bibnamefont
  {Suzuki}},\ }\bibfield  {title} {\bibinfo {title} {Nuisance parameter problem
  in quantum estimation theory: Tradeoff relation and qubit examples},\ }\href
  {https://doi.org/10.1088/1751-8121/ab8672} {\bibfield  {journal} {\bibinfo
  {journal} {J. Phys. A}\ }\textbf {\bibinfo {volume} {53}},\ \bibinfo {pages}
  {264001} (\bibinfo {year} {2020})}\BibitemShut {NoStop}%
\bibitem [{\citenamefont {Giovannini}\ \emph {et~al.}(2015)\citenamefont
  {Giovannini}, \citenamefont {Romero}, \citenamefont {Poto{\v c}ek},
  \citenamefont {Ferenczi}, \citenamefont {Speirits}, \citenamefont {Barnett},
  \citenamefont {Faccio},\ and\ \citenamefont {Padgett}}]{Giovannini857}%
  \BibitemOpen
  \bibfield  {author} {\bibinfo {author} {\bibfnamefont {D.}~\bibnamefont
  {Giovannini}}, \bibinfo {author} {\bibfnamefont {J.}~\bibnamefont {Romero}},
  \bibinfo {author} {\bibfnamefont {V.}~\bibnamefont {Poto{\v c}ek}}, \bibinfo
  {author} {\bibfnamefont {G.}~\bibnamefont {Ferenczi}}, \bibinfo {author}
  {\bibfnamefont {F.}~\bibnamefont {Speirits}}, \bibinfo {author}
  {\bibfnamefont {S.~M.}\ \bibnamefont {Barnett}}, \bibinfo {author}
  {\bibfnamefont {D.}~\bibnamefont {Faccio}},\ and\ \bibinfo {author}
  {\bibfnamefont {M.~J.}\ \bibnamefont {Padgett}},\ }\bibfield  {title}
  {\bibinfo {title} {Spatially structured photons that travel in free space
  slower than the speed of light},\ }\href
  {https://doi.org/10.1126/science.aaa3035} {\bibfield  {journal} {\bibinfo
  {journal} {Science}\ }\textbf {\bibinfo {volume} {347}},\ \bibinfo {pages}
  {857} (\bibinfo {year} {2015})}\BibitemShut {NoStop}%
\bibitem [{\citenamefont {Chen}\ \emph {et~al.}(2019)\citenamefont {Chen},
  \citenamefont {Fink}, \citenamefont {Steinlechner}, \citenamefont {Torres},\
  and\ \citenamefont {Ursin}}]{Chen2019g}%
  \BibitemOpen
  \bibfield  {author} {\bibinfo {author} {\bibfnamefont {Y.}~\bibnamefont
  {Chen}}, \bibinfo {author} {\bibfnamefont {M.}~\bibnamefont {Fink}}, \bibinfo
  {author} {\bibfnamefont {F.}~\bibnamefont {Steinlechner}}, \bibinfo {author}
  {\bibfnamefont {J.~P.}\ \bibnamefont {Torres}},\ and\ \bibinfo {author}
  {\bibfnamefont {R.}~\bibnamefont {Ursin}},\ }\bibfield  {title} {\bibinfo
  {title} {Hong-{{Ou}}-{{Mandel}} interferometry on a biphoton beat note},\
  }\href {https://doi.org/10.1038/s41534-019-0161-z} {\bibfield  {journal}
  {\bibinfo  {journal} {npj Quantum Inf.}\ }\textbf {\bibinfo {volume} {5}},\
  \bibinfo {pages} {43} (\bibinfo {year} {2019})}\BibitemShut {NoStop}%
\bibitem [{\citenamefont {Scott}\ \emph {et~al.}(2020)\citenamefont {Scott},
  \citenamefont {Branford}, \citenamefont {Westerberg}, \citenamefont {Leach},\
  and\ \citenamefont {Gauger}}]{Scott2020}%
  \BibitemOpen
  \bibfield  {author} {\bibinfo {author} {\bibfnamefont {H.}~\bibnamefont
  {Scott}}, \bibinfo {author} {\bibfnamefont {D.}~\bibnamefont {Branford}},
  \bibinfo {author} {\bibfnamefont {N.}~\bibnamefont {Westerberg}}, \bibinfo
  {author} {\bibfnamefont {J.}~\bibnamefont {Leach}},\ and\ \bibinfo {author}
  {\bibfnamefont {E.~M.}\ \bibnamefont {Gauger}},\ }\bibfield  {title}
  {\bibinfo {title} {Beyond coincidence in {{Hong}}-{{Ou}}-{{Mandel}}
  interferometry},\ }\href {https://doi.org/10.1103/PhysRevA.102.033714}
  {\bibfield  {journal} {\bibinfo  {journal} {Phys. Rev. A}\ }\textbf {\bibinfo
  {volume} {102}},\ \bibinfo {pages} {033714} (\bibinfo {year}
  {2020})}\BibitemShut {NoStop}%
\bibitem [{Note1()}]{Note1}%
  \BibitemOpen
  \bibinfo {note} {Strictly speaking, these are already written in
  infinitesimal form, as they would appear inside an integral, but the measure
  is actually defined on any subset of the detector space.}\BibitemShut {Stop}%
\bibitem [{Note2()}]{Note2}%
  \BibitemOpen
  \bibinfo {note} {Notice that our definition of the HG functions differs from
  the usual eigenfunctions of the Fourier transform, as we have used the
  multiplying factor $e^{-\xi ^2\omega ^2}$ instead of $e^{-\xi ^2\omega
  ^2/2}$.}\BibitemShut {Stop}%
\bibitem [{\citenamefont {Ushakov}(2011)}]{Ushakov2011}%
  \BibitemOpen
  \bibfield  {author} {\bibinfo {author} {\bibfnamefont {N.~G.}\ \bibnamefont
  {Ushakov}},\ }\href
  {https://www.degruyter.com/document/doi/10.1515/9783110935981/html} {\emph
  {\bibinfo {title} {Selected {{Topics}} in {{Characteristic Functions}}}}}\
  (\bibinfo  {publisher} {{De Gruyter}},\ \bibinfo {year} {2011})\BibitemShut
  {NoStop}%
\bibitem [{\citenamefont {Mathias}(1923)}]{Mathias1923}%
  \BibitemOpen
  \bibfield  {author} {\bibinfo {author} {\bibfnamefont {M.}~\bibnamefont
  {Mathias}},\ }\bibfield  {title} {\bibinfo {title} {{\"Uber positive
  Fourier-Integrale}},\ }\href {https://doi.org/10.1007/BF01175675} {\bibfield
  {journal} {\bibinfo  {journal} {Math. Z.}\ }\textbf {\bibinfo {volume}
  {16}},\ \bibinfo {pages} {103} (\bibinfo {year} {1923})}\BibitemShut
  {NoStop}%
\bibitem [{\citenamefont {Wang}(2006)}]{Wang2006a}%
  \BibitemOpen
  \bibfield  {author} {\bibinfo {author} {\bibfnamefont {K.}~\bibnamefont
  {Wang}},\ }\bibfield  {title} {\bibinfo {title} {Quantum theory of two-photon
  wavepacket interference in a beamsplitter},\ }\href
  {https://doi.org/10.1088/0953-4075/39/18/R01} {\bibfield  {journal} {\bibinfo
   {journal} {J. Phys. B}\ }\textbf {\bibinfo {volume} {39}},\ \bibinfo {pages}
  {R293} (\bibinfo {year} {2006})}\BibitemShut {NoStop}%
\bibitem [{\citenamefont {Gianani}\ \emph {et~al.}(2021)\citenamefont
  {Gianani}, \citenamefont {Albarelli}, \citenamefont {Cimini},\ and\
  \citenamefont {Barbieri}}]{Gianani2020}%
  \BibitemOpen
  \bibfield  {author} {\bibinfo {author} {\bibfnamefont {I.}~\bibnamefont
  {Gianani}}, \bibinfo {author} {\bibfnamefont {F.}~\bibnamefont {Albarelli}},
  \bibinfo {author} {\bibfnamefont {V.}~\bibnamefont {Cimini}},\ and\ \bibinfo
  {author} {\bibfnamefont {M.}~\bibnamefont {Barbieri}},\ }\bibfield  {title}
  {\bibinfo {title} {Experimental function estimation from quantum phase
  measurements},\ }\href {https://doi.org/10.1103/PhysRevA.103.042602}
  {\bibfield  {journal} {\bibinfo  {journal} {Phys. Rev. A}\ }\textbf {\bibinfo
  {volume} {103}},\ \bibinfo {pages} {042602} (\bibinfo {year}
  {2021})}\BibitemShut {NoStop}%
\bibitem [{\citenamefont {Harnchaiwat}\ \emph {et~al.}(2020)\citenamefont
  {Harnchaiwat}, \citenamefont {Zhu}, \citenamefont {Westerberg}, \citenamefont
  {Gauger},\ and\ \citenamefont {Leach}}]{Harnchaiwat2020}%
  \BibitemOpen
  \bibfield  {author} {\bibinfo {author} {\bibfnamefont {N.}~\bibnamefont
  {Harnchaiwat}}, \bibinfo {author} {\bibfnamefont {F.}~\bibnamefont {Zhu}},
  \bibinfo {author} {\bibfnamefont {N.}~\bibnamefont {Westerberg}}, \bibinfo
  {author} {\bibfnamefont {E.}~\bibnamefont {Gauger}},\ and\ \bibinfo {author}
  {\bibfnamefont {J.}~\bibnamefont {Leach}},\ }\bibfield  {title} {\bibinfo
  {title} {Tracking the polarisation state of light via
  {{Hong}}-{{Ou}}-{{Mandel}} interferometry},\ }\href
  {https://doi.org/10.1364/OE.382622} {\bibfield  {journal} {\bibinfo
  {journal} {Opt. Express}\ }\textbf {\bibinfo {volume} {28}},\ \bibinfo
  {pages} {2210} (\bibinfo {year} {2020})}\BibitemShut {NoStop}%
\bibitem [{\citenamefont {Scott}\ \emph {et~al.}(2021)\citenamefont {Scott},
  \citenamefont {Branford}, \citenamefont {Westerberg}, \citenamefont {Leach},\
  and\ \citenamefont {Gauger}}]{Scott2021}%
  \BibitemOpen
  \bibfield  {author} {\bibinfo {author} {\bibfnamefont {H.}~\bibnamefont
  {Scott}}, \bibinfo {author} {\bibfnamefont {D.}~\bibnamefont {Branford}},
  \bibinfo {author} {\bibfnamefont {N.}~\bibnamefont {Westerberg}}, \bibinfo
  {author} {\bibfnamefont {J.}~\bibnamefont {Leach}},\ and\ \bibinfo {author}
  {\bibfnamefont {E.~M.}\ \bibnamefont {Gauger}},\ }\href@noop {} {\bibinfo
  {title} {Noise limits on two-photon interferometric sensing}} (\bibinfo
  {year} {2021}),\ \Eprint {https://arxiv.org/abs/2106.13671} {arXiv:2106.13671
  [quant-ph]} \BibitemShut {NoStop}%
\bibitem [{\citenamefont {Ndagano}\ \emph {et~al.}(2021)\citenamefont
  {Ndagano}, \citenamefont {Defienne}, \citenamefont {Branford}, \citenamefont
  {Shah}, \citenamefont {Lyons}, \citenamefont {Westerberg}, \citenamefont
  {Gauger},\ and\ \citenamefont {Faccio}}]{Ndagano2021}%
  \BibitemOpen
  \bibfield  {author} {\bibinfo {author} {\bibfnamefont {B.}~\bibnamefont
  {Ndagano}}, \bibinfo {author} {\bibfnamefont {H.}~\bibnamefont {Defienne}},
  \bibinfo {author} {\bibfnamefont {D.}~\bibnamefont {Branford}}, \bibinfo
  {author} {\bibfnamefont {Y.~D.}\ \bibnamefont {Shah}}, \bibinfo {author}
  {\bibfnamefont {A.}~\bibnamefont {Lyons}}, \bibinfo {author} {\bibfnamefont
  {N.}~\bibnamefont {Westerberg}}, \bibinfo {author} {\bibfnamefont {E.~M.}\
  \bibnamefont {Gauger}},\ and\ \bibinfo {author} {\bibfnamefont
  {D.}~\bibnamefont {Faccio}},\ }\href@noop {} {\bibinfo {title}
  {Hong-{{Ou}}-{{Mandel}} microscopy}} (\bibinfo {year} {2021}),\ \Eprint
  {https://arxiv.org/abs/2108.05346} {arXiv:2108.05346 [quant-ph]} \BibitemShut
  {NoStop}%
\bibitem [{\citenamefont {Sbroscia}\ \emph {et~al.}(2018)\citenamefont
  {Sbroscia}, \citenamefont {Gianani}, \citenamefont {Roccia}, \citenamefont
  {Cimini}, \citenamefont {Mancino}, \citenamefont {Aloe},\ and\ \citenamefont
  {Barbieri}}]{sbroscia}%
  \BibitemOpen
  \bibfield  {author} {\bibinfo {author} {\bibfnamefont {M.}~\bibnamefont
  {Sbroscia}}, \bibinfo {author} {\bibfnamefont {I.}~\bibnamefont {Gianani}},
  \bibinfo {author} {\bibfnamefont {E.}~\bibnamefont {Roccia}}, \bibinfo
  {author} {\bibfnamefont {V.}~\bibnamefont {Cimini}}, \bibinfo {author}
  {\bibfnamefont {L.}~\bibnamefont {Mancino}}, \bibinfo {author} {\bibfnamefont
  {P.}~\bibnamefont {Aloe}},\ and\ \bibinfo {author} {\bibfnamefont
  {M.}~\bibnamefont {Barbieri}},\ }\bibfield  {title} {\bibinfo {title}
  {Assessing frequency correlation through a distinguishability measurement},\
  }\href {https://doi.org/10.1364/OL.43.004045} {\bibfield  {journal} {\bibinfo
   {journal} {Opt. Lett.}\ }\textbf {\bibinfo {volume} {43}},\ \bibinfo {pages}
  {4045} (\bibinfo {year} {2018})}\BibitemShut {NoStop}%
\bibitem [{\citenamefont {Flamini}\ \emph {et~al.}(2015)\citenamefont
  {Flamini}, \citenamefont {Magrini}, \citenamefont {Rab}, \citenamefont
  {Spagnolo}, \citenamefont {D'Ambrosio}, \citenamefont {Mataloni},
  \citenamefont {Sciarrino}, \citenamefont {Zandrini}, \citenamefont {Crespi},
  \citenamefont {Ramponi},\ and\ \citenamefont {Osellame}}]{flamini}%
  \BibitemOpen
  \bibfield  {author} {\bibinfo {author} {\bibfnamefont {F.}~\bibnamefont
  {Flamini}}, \bibinfo {author} {\bibfnamefont {L.}~\bibnamefont {Magrini}},
  \bibinfo {author} {\bibfnamefont {A.~S.}\ \bibnamefont {Rab}}, \bibinfo
  {author} {\bibfnamefont {N.}~\bibnamefont {Spagnolo}}, \bibinfo {author}
  {\bibfnamefont {V.}~\bibnamefont {D'Ambrosio}}, \bibinfo {author}
  {\bibfnamefont {P.}~\bibnamefont {Mataloni}}, \bibinfo {author}
  {\bibfnamefont {F.}~\bibnamefont {Sciarrino}}, \bibinfo {author}
  {\bibfnamefont {T.}~\bibnamefont {Zandrini}}, \bibinfo {author}
  {\bibfnamefont {A.}~\bibnamefont {Crespi}}, \bibinfo {author} {\bibfnamefont
  {R.}~\bibnamefont {Ramponi}},\ and\ \bibinfo {author} {\bibfnamefont
  {R.}~\bibnamefont {Osellame}},\ }\bibfield  {title} {\bibinfo {title}
  {Thermally reconfigurable quantum photonic circuits at telecom wavelength by
  femtosecond laser micromachining},\ }\href
  {https://doi.org/10.1038/lsa.2015.127} {\bibfield  {journal} {\bibinfo
  {journal} {Light Sci. Appl.}\ }\textbf {\bibinfo {volume} {4}},\ \bibinfo
  {pages} {e354} (\bibinfo {year} {2015})}\BibitemShut {NoStop}%
\end{thebibliography}%
    
\end{document}